\documentclass[aps,twocolumn,prc,showpacs]{revtex4}
\usepackage{mathrsfs}
\usepackage{longtable,lscape}
\usepackage{txfonts}
\usepackage{amssymb}
\usepackage{indentfirst}
\usepackage{graphicx,,booktabs}
\usepackage{color}
\usepackage{amssymb}
\usepackage{epsfig}
\newcommand{\vsig}{\mbox{\boldmath$\sigma$\unboldmath}}
\newcommand{\veps}{\mbox{\boldmath$\epsilon$\unboldmath}}

\begin{document}

\title{Neutral pion photoproduction on the nucleon in a chiral quark model}
\author{
Li-Ye Xiao$^{1}$, Xu Cao$^{2}$ and Xian-Hui Zhong$^{1}$ \footnote
{E-mail: zhongxh@hunnu.edu.cn} } \affiliation{ 1)  Department of
Physics, Hunan Normal University, and Key Laboratory of
Low-Dimensional Quantum Structures and Quantum Control of Ministry
of Education, Changsha 410081, China }

\affiliation{ 2) Institute of Modern Physics, Chinese Academy of
Sciences, Lanzhou 730000, China}

\begin{abstract}
A chiral quark-model approach is adopted to study the $\gamma
p\rightarrow \pi^0p$ and $\gamma n\rightarrow \pi^0 n$ reactions.
Good descriptions of the total and differential cross sections and
single-polarization observables are obtained from the pion
production threshold up to the second resonance region. It is found
that (i) the $n=0$ shell resonance $\Delta(1232)P_{33}$, the $n=1$
shell resonances $N(1535)S_{11}$ and $N(1520)D_{13}$, and the $n=2$
shell resonance $N(1720)P_{13}$ play crucial roles in these two
processes. They are responsible for the first, second and third bump
structures in the cross sections, respectively. (ii) Furthermore,
obvious evidences of $N(1650)S_{11}$ and $\Delta(1620)S_{31}$ are
also found in the reactions. They notably affect the cross sections
and the polarization observables from the second resonance region to
the third resonance region. (iii) The $u$-channel background plays a
crucial role in the reactions. It has strong interferences with the
$s$-channel resonances. (iv) The $t$-channel background seems to be
needed in the reactions. Including the $t$-channel vector-meson
exchange contribution, the descriptions in the energy region
$E_\gamma= 600\sim 900$ MeV are improved significantly. The helicity
amplitudes of the main resonances, $\Delta(1232)P_{33}$,
$N(1535)S_{11}$, $N(1520)D_{13}$, $N(1720)P_{13}$, $N(1650)S_{11}$,
and $\Delta(1620)S_{31}$, are extracted and compared with the
results from other groups.

\end{abstract}
 \maketitle

\section{Introduction}

Understanding of the baryon spectrum and searching for the missing
nucleon resonances and new exotic states are hot topics in hadronic
physics~\cite{Klempt:2009pi}. Photoproduction of mesons is an ideal
tool for the study of nucleon and $\Delta(1232)$ spectroscopies in
experiments~\cite{Krusche:2003ik}. Neutral pion photoproduction
reactions are of special interest because the neutral pions do not
couple directly to photons so that nonresonant background
contributions are suppressed (i.e., no contact term
contribution)~\cite{Dieterle:2014blj}. In the past few years, great
progress has been achieved in experiments studying of the $\gamma
p\rightarrow \pi^0 p$ reaction at JLab, CB-ELSA, MAMI, and GRAAL.
These experimental groups have carried out precise measurements of
the differential cross sections and single-polarization observables
with a large solid angle coverage and a wide photon energy
range~\cite{Fuchs:1996ja,Beck:2006ye,Sparks:2010vb,Crede:2011dc,vanPee:2007tw,Elsner:2008sn,
Adamian:2000yi,Dugger:2007bt,Bartalini:2005wx,Hartmann:2014mya,Bartholomy:2004uz,Anisovich:2005tf}.
Recently they also have finished some measurements of the
double-polarization
observables~\cite{Hartmann:2014mya,Thiel:2012yj,Sikora:2013vfa,Gottschall:2013uha}.
Furthermore, in recent years significant progress has been achieved
in experiments measuring the $\gamma n\rightarrow \pi^0 n$ reaction
as well. In 2009, some measurements of the beam asymmetries for the
$\gamma n\rightarrow \pi^0 n$ process were obtained by the GRAAL
experiment in the second and third resonances
region~\cite{DiSalvo:2009zz}. Very recently, the quasi-free
differential and total cross sections in the second and third
resonances region for this reaction were also measured by the
Crystal Ball/TAPS experiment at MAMI~\cite{Dieterle:2014blj}. Thus,
improvement of the experimental situations gives us a good
opportunity to study the excitation spectroscopies of the nucleon
and $\Delta(1232)$.

Stimulated by these new measurements, many partial-wave analysis
groups, such as
BnGa~\cite{Anisovich:2009zy,Anisovich:2011fc,Anisovich:2013jya},
SAID~\cite{Chen:2012yv,Workman:2011vb,Workman:2012jf},
MAID~\cite{Drechsel:2007if}, Kent~\cite{Shrestha:2012ep},
J\"{u}lich~\cite{Huang:2011as,Ronchen:2014cna} and
ANL-Osaka~\cite{Kamano:2013iva}, have updated their analysis in
recent years. For the $\gamma p\rightarrow \pi^0 p$ reaction, good
descriptions of the data up to the second and third resonances
region have been obtained by different groups. However, the
explanations of the reaction data and the extracted resonance
properties from the reaction still exhibit strong model
dependencies. For example, the $\gamma p$ couplings for some
well-established resonances, such as $N(1535)S_{11}$,
$N(1650)S_{11}$ and $N(1520)D_{13}$, extracted by various groups
differ rather notably from each other. For the $\gamma n\rightarrow
\pi^0 n$ reaction, consistent predictions from different approaches
can only be obtained in the first resonance
region~\cite{Dieterle:2014blj}. Because of the lack of data, the
predictions from different models in the second and third resonances
region are very different. Fortunately, in this energy region some
new measurements of the cross section for the $\gamma n\rightarrow
\pi^0 n$ reaction at MAMI~\cite{Dieterle:2014blj} were reported
about one year ago.

These new data for the $\gamma n\rightarrow \pi^0 n$ reaction not
only provide us a good opportunity to extract more knowledge of the
neutron resonances, but also shed light on the puzzle of the narrow
structure around $W=1.68$ GeV observed in the excitation function of
$\eta$ production off quasi-free neutrons by several experimental
groups~\cite{Kuznetsov:2007gr, Jaegle:2008ux, Jaegle:2011sw}. This
narrow structure has been listed by the Particle Data Group (PDG) as
a new nucleon resonance $N(1685)$~\cite{Agashe:2014kda}. However,
many controversial explanations about this narrow structure, such as
the $N(1650)S_{11}$ and $N(1710)P_{11}$ coupled-channel effects,
interference effects between $N(1650)S_{11}$, $N(1710)P_{11}$ and
$N(1720)P_{13}$, and effects from strangeness threshold openings,
can be found in the
literature~\cite{Shyam:2008fr,Shklyar:2006xw,Doring:2009qr}. In our
quark model study, we find that the narrow structure around $W=
1.68$ GeV can be explained by the constructive interferences between
$N(1535)S_{11}$ and $N(1650)S_{11}$~\cite{Zhong:2011ti}. Our
conclusion is consistent with the analysis from the BnGa
group~\cite{Anisovich:2015tla,Anisovich:2008wd}. It should be
mentioned that, the $\gamma n$ coupling for $N(1650)S_{11}$
extracted by us and BnGa group has a positive sign, which is
opposite to that of PDG~\cite{Agashe:2014kda}. Now, two questions
arise naturally: (i) Can some clues about the controversially
discussed $N(1685)$ be found in the $\gamma n\rightarrow \pi^0 n$
reaction? (ii) Are the properties of $N(1535)S_{11}$ and
$N(1650)S_{11}$ extracted from the $\eta N$ channel consistent with
those extracted from the $\pi^0 N$ channel? To better understand
these questions, a systematic analysis of the recent data for the
neutral pion production off nucleons is urgently needed.

In this work, we carry out a combined study of the $\gamma
p\rightarrow \pi^0 p$ and $\gamma n\rightarrow \pi^0 n$ reactions in
a chiral quark model. By systematically analyzing the new data for
neutral pion photoproduction on the nucleons, we attempt to uncover
some puzzles existing in the photoproduction reactions and obtain a
better understanding of the excitation spectra of the nucleon and
$\Delta(1232)$. It should be mentioned that there are interesting
differences between $\gamma p\to \pi^0 p$ and  $\gamma n\to \pi^0
n$. In the $\gamma p$ reactions, contributions from the nucleon
resonances of representation $[70,^48]$ will be suppressed by the
Moorhouse selection rule~\cite{Moorhouse:1966jn,Zhao:2006an}. In
contrast, all the octet states can contribute to the $\gamma n$
reactions. In other words, more states will be present in the
$\gamma n$ reactions. Therefore, by studying neutral pion
photoproduction on nucleons, we expect that the role played by
intermediate baryon resonances can be highlighted.

In the chiral quark model, an effective chiral Lagrangian is
introduced to account for the quark-pseudoscalar-meson coupling.
Since the quark-meson coupling is invariant under the chiral
transformation, some of the low-energy properties of QCD are
retained. The chiral quark model has been well developed and widely
applied to meson photoproduction reactions
~\cite{Zhong:2011ti,Zhong:2011ht,Li:1994cy,Li:1995si,Li:1997gd,
Zhao:2002id,Zhao:1998fn,Zhao:2001jw,Zhao:1998rt,
Li:1998ni,Saghai:2001yd,He:2008ty,He:2009zzi}. Recently, this model
has been successfully extended to $\pi N$ and $KN$ reactions as
well~\cite{Zhong:2007fx,Zhong:2008km,Zhong:2013oqa,Xiao:2013hca}.


The paper is organized as follows. In Sec.~\ref{fram}, a brief
review of the chiral quark model approach is given. The numerical
results are presented and discussed in Sec.~\ref{discuss}. Finally,
a summary is given in Sec.~\ref{sum}.

\section{The model}\label{fram}

In this section, we give a brief review of the chiral quark model.
In this model, the $s$- and $u$-channel transition amplitudes are
determined by~\cite{Li:1997gd,Li:1995si}
\begin{eqnarray}
\mathcal{M}_{s}=\sum_j\langle N_f |H_{m} |N_j\rangle\langle N_j
|\frac{1}{E_i+\omega_\gamma-E_j}H_{e}|N_i\rangle,\\
\mathcal{M}_{u}=\sum_j\langle N_f |H_{e }
\frac{1}{E_i-\omega_m-E_j}|N_j\rangle\langle N_j | H_{m }
|N_i\rangle,
\end{eqnarray}
where $H_{m}$ and $H_e$ stand for the quark-pseudoscalar-meson and
electromagnetic couplings at the tree level, respectively.  They are
described by~\cite{Li:1995si,Li:1997gd,Zhao:2002id}
\begin{eqnarray}
H_m&=&\sum_j
\frac{1}{f_m}\bar{\psi}_j\gamma^{j}_{\mu}\gamma^{j}_{5}\psi_j\vec{\tau}\cdot\partial^{\mu}\vec{\phi}_m,\label{coup}\\
H_e&=&-\sum_je_j\gamma^{j}_{\mu}A^{\mu}(\mathbf{k},\mathbf{r})\label{coup1},
\end{eqnarray}
where $\psi_j$ represents the $j$-th quark field in a hadron,
$\phi_m$ is the field of the pseudoscalar-meson octet,  and $f_m$ is
the meson's decay constant. The $\omega_\gamma$ is the energy of the
incoming photons. The $|N_i\rangle$, $|N_j\rangle$ and $|N_f\rangle$
stand for the initial, intermediate and final states, respectively,
and their corresponding energies are $E_i$, $E_j$ and $E_f$, which
are the eigenvalues of the non-relativistic constituent quark model
Hamiltonian
$\hat{H}$~\cite{Isgur:1978xj,Isgur:1977ef,Capstick:1986bm}. The $s$-
and $u$-channel transition amplitudes have been worked out in the
harmonic oscillator basis in
Refs.~\cite{Li:1995si,Li:1997gd,Zhao:2002id}.

The $t$-channel contributions of vector meson exchange are included
in this work. The effective Lagrangians for the vector meson
exchange for the $\gamma\pi V$ and $Vqq$ couplings are adopted
as~\cite{Zhao:2002id}
\begin{eqnarray}
\mathcal{L}_{\gamma\pi
V}&=&e \frac{g_{V\pi \gamma}}{m_\pi}\varepsilon_{\alpha\beta\gamma\delta}\partial^{\alpha}A^{\beta}\partial^{\gamma}V^{\delta}\pi,\\
\mathcal{L}_{Vqq}&=&g_{Vqq}\bar{\psi}_j(\gamma_{\mu}+\frac{\kappa_q}{2m_q}\sigma_{\mu\nu}\partial^{\nu})V^{\mu}\psi_j,
\end{eqnarray}
where $A$ and $V$ denote the photon and vector-meson fields,
respectively; $\pi$ stands for the $\pi$-meson field; $g_{V\pi
\gamma}$ and $g_{Vqq}$ are the coupling constants. The $t$-channel
transition amplitude has been given in the harmonic oscillator basis
in Refs.~\cite{Zhao:2002id}.

It should be remarked that the amplitudes in terms of the harmonic
oscillator principle quantum number $n$ are the sum of a set of
SU(6) multiplets with the same $n$. To obtain the contributions of
individual resonances, we need to separate out the
single-resonance-excitation amplitudes within each principle number
$n$ in the $s$-channel. Taking into account the width effects of the
resonances, the resonance transition amplitudes of the $s$-channel
can be generally expressed as \cite{Li:1997gd}
\begin{eqnarray}
\mathcal{M}^s_R=\frac{2M_R}{s-M^2_R+iM_R
\Gamma_R}\mathcal{O}_Re^{-(\textbf{k}^2+\textbf{q}^2)/6\alpha^2},
\label{stt}
\end{eqnarray}
where $\sqrt{s}=E_i+\omega_\gamma$ is the total energy of the
system, $\alpha$ is the harmonic oscillator strength, $M_R$ is the
mass of the $s$-channel resonance with a width $\Gamma_R$, and
$\mathcal{O}_R$ is the separated operators for individual resonances
in the $s$-channel. In the Chew-Goldberger-Low-Nambu (CGLN)
parameterization, the transition amplitude can be written in a
standard form~\cite{Chew:1957tf}:
\begin{eqnarray}
\mathcal{O}_R&=&i f^R_1 \vsig \cdot \veps+f^R_2 \frac{(\vsig \cdot
\mathbf{q})\vsig\cdot (\mathbf{k}\times
\veps)}{|\mathbf{q}||\mathbf{k}|}\nonumber\\
&& +if^R_3\frac{(\vsig \cdot \mathbf{k}) (\mathbf{q}\cdot
\veps)}{|\mathbf{q}||\mathbf{k}|}+if^R_4\frac{(\vsig \cdot
\mathbf{q}) (\mathbf{q}\cdot \veps)}{|\mathbf{q}|^2},
\end{eqnarray}
where $\vsig$ is the spin operator of the nucleon, $\veps$ is the
polarization vector of the photon, and $\mathbf{k}$ and $\mathbf{q}$
are incoming photon and outgoing meson momenta, respectively. In the
SU(6)$\otimes$O(3) symmetry limit, we have extracted the CGLN
amplitudes for the $s$-channel resonances in the $n\leq2$ shell for
the $\gamma p\rightarrow \pi^0 p$ and $\gamma n \rightarrow \pi^0 n$
processes, which have been listed in Tables~\ref{CGLN1}
and~\ref{CGLN2}, respectively. Comparing the CGLN amplitudes of
different resonances with each other, one can easily find which
states are the main contributors to the reactions in the
SU(6)$\otimes$O(3) symmetry limit.

\begin{widetext}
\begin{center}
\begin{table}[ht]
\caption{The CGLN amplitudes of $s$-channel resonances in the
$n\leq2$ shell for the $\gamma p\rightarrow \pi^0 p$ process in the
SU(6)$\otimes$O(3) symmetry limit. We have defined
$A\equiv-(\frac{\omega_m}{E_f+M_N}+1)|\mathbf{q}|$,
$B\equiv\frac{\omega_m}{\mu_q}+\frac{2|\mathbf{q}|}{3\alpha^2}A$,
$C\equiv\frac{\omega_m}{\mu_q}+\frac{|\mathbf{q}|}{\alpha^2}A$,
$D\equiv\frac{\omega_m}{\mu_q}+\frac{2|\mathbf{q}|}{5\alpha^2}A$,
$x\equiv\frac{|\mathbf{k}| |\mathbf{q}|}{3\alpha^2}$, $P_l'(z)\equiv
\frac{\partial P_l(z)}{\partial z}$, $P_l''(z)\equiv\frac{\partial^2
P_l(z)}{\partial z^2}$. The $\omega_\gamma$, $\omega_m$ and $E_f$
stand for the energies of the incoming photon, outgoing meson and
final nucleon, respectively. The $m_q$ is the constituent $u$ or $d$
quark mass. $1/\mu_q$ is a factor defined by $1/\mu_q=2/m_q$.
$P_l(z)$ is the Legendre function with $z=\cos\theta$.}
\label{CGLN1}
\begin{tabular}{|c|c|c|c|c|c|c|c|c|c| }\hline\hline
resonance &$[N_6,^{2S+1}N_3,n,l]$  & $f^R_1$ \ \ & $f^R_2$ & $f^R_3$ & $f^R_4$  \\
\hline $N(938)P_{11}$&$[56,^28,0,0]$&0& $+i\frac{5\sqrt{2}}{2}\frac{k}{6m_q}A$ &0 &0\\
\hline $\Delta(1232)P_{33}$&$[56,^410,0,0]$&
$+i\frac{4\sqrt{2}}{3}\frac{k}{6m_q}A P_2'(z)$ &
$+i\frac{8\sqrt{2}}{3}\frac{k}{6m_q}A$
&$-i\frac{4\sqrt{2}}{3}\frac{k}{6m_q}A
P_2''(z)$ &  0\\
       \hline
$N(1535)S_{11}$&$[70,^28,1,1]$&
$-i\frac{\sqrt{2}}{18}k(1+\frac{k}{2m_q})B$
&0 &0 &  0\\
\hline $\Delta(1620)S_{31}$&$[70,^210,1,1]$&
$+i\frac{\sqrt{2}}{36}k(1-\frac{k}{6m_q})B$
&0 &0 &  0\\
\hline $N(1520)D_{13}$&$[70,^28,1,1]$&
$+i\frac{\sqrt{2}}{27}k(1+\frac{k}{2m_q})\frac{|\mathbf{q}|}{\alpha^2}A$
&$+i\frac{\sqrt{2}}{54}k\frac{k}{m_q}\frac{|\mathbf{q}|}{\alpha^2}AP_2'(z)$
&0 &
$-i\frac{\sqrt{2}}{27}k\frac{|\mathbf{q}|}{\alpha^2}AP_2''(z)$\\
\hline $\Delta(1700)D_{33}$&$[70,^210,1,1]$&
$-i\frac{\sqrt{2}}{54}k(1-\frac{k}{6m_q})\frac{|\mathbf{q}|}{\alpha^2}A$
&$+i\frac{\sqrt{2}}{54}k\frac{k}{6m_q}\frac{|\mathbf{q}|}{\alpha^2}AP_2'(z)$
&0 &
$+i\frac{\sqrt{2}}{54}k\frac{|\mathbf{q}|}{\alpha^2}AP_2''(z)$\\
\hline $N(1440)P_{11}$&$[56,^28,2,0]$& 0
&$+i\frac{11\sqrt{2}}{36\times18}\frac{15}{19}k\frac{k}{m_q}Cx$ &0 &
0\\
\hline $N(1710)P_{11}$&$[70,^28,2,0]$& 0
&$+i\frac{11\sqrt{2}}{36\times18}\frac{6}{19}k\frac{k}{m_q}Cx$ &0 &
0\\
\hline $\Delta(1750)P_{31}$&$[70,^210,2,0]$& 0
&$-i\frac{11\sqrt{2}}{36\times18}\frac{2}{19}k\frac{k}{m_q}Cx$ &0 &
0\\
\hline
$N(1720)P_{13}$&$[56,^28,2,2]$&$-i\frac{\sqrt{2}}{90}\frac{25}{12}k(1+\frac{k}{2m_q})DP_2'(z)x$
&$-i\frac{\sqrt{2}}{90}\frac{25}{12}k\frac{k}{2m_q}Dx$
&$-i\frac{\sqrt{2}}{90}\frac{25}{12}kDP_2''(z)x$&
0\\
\hline
$N(1900)P_{13}$&$[70,^28,2,2]$&$-i\frac{\sqrt{2}}{90}\frac{10}{12}k(1+\frac{k}{2m_q})DP_2'(z)x$
&$-i\frac{\sqrt{2}}{90}\frac{10}{12}k\frac{k}{2m_q}Dx$
&$-i\frac{\sqrt{2}}{90}\frac{10}{12}kDP_2''(z)x$&
0\\
\hline
$\Delta(1985?)P_{33}$&$[70,^210,2,2]$&$+i\frac{\sqrt{2}}{90}\frac{5}{12}k(1-\frac{k}{6m_q})DP_2'(z)x$
&$-i\frac{\sqrt{2}}{90}\frac{5}{12}k\frac{k}{6m_q}Dx$
&$+i\frac{\sqrt{2}}{90}\frac{5}{12}kDP_2''(z)x$&
0\\
\hline $\Delta(1920)P_{33}$&$[56,^410,2,2]$&0
&$-i\frac{\sqrt{2}}{90}\frac{10}{9}k\frac{k}{2m_q}Dx$
&$+i\frac{\sqrt{2}}{90}\frac{10}{9}k\frac{k}{2m_q}DP_2''(z)x$&
0\\
\hline
$\Delta(1600)P_{33}$&$[56,^410,2,0]$&$+i\frac{\sqrt{2}}{90}\frac{10}{9}k\frac{k}{2m_q}CP_2'(z)x$
&$+i\frac{\sqrt{2}}{90}\frac{20}{9}k\frac{k}{2m_q}Cx$
&$-i\frac{\sqrt{2}}{90}\frac{10}{9}k\frac{k}{2m_q}CP_2''(z)x$&
0\\
\hline
$\Delta(1905)F_{35}$&$[56,^410,2,2]$&$+i\frac{2\sqrt{2}}{3}\frac{5k}{630m_q}AP_2'(z)x^2$
&$+i\frac{2\sqrt{2}}{3}\frac{2k}{630m_q}AP_3'(z)x^2$
&$+i\frac{2\sqrt{2}}{3}\frac{3k}{630m_q}AP_2''(z)x^2$&
$-i\frac{2\sqrt{2}}{3}\frac{3k}{630m_q}AP_3''(z)x^2$\\
\hline
$\Delta(?)F_{35}$&$[70,^210,2,2]$&$-i\frac{\sqrt{2}}{180}(1-\frac{k}{6m_q})AP_2'(z)x^2$
&$+i\frac{\sqrt{2}}{180}\frac{k}{6m_q}AP_3'(z)x^2$
&$-i\frac{\sqrt{2}}{180}AP_2''(z)x^2$&
$+i\frac{\sqrt{2}}{180}AP_3''(z)x^2$\\
\hline
$N(1680)F_{15}$&$[56,^28,2,2]$&$+i\frac{5\sqrt{2}}{180}(1+\frac{k}{2m_q})AP_2'(z)x^2$
&$+i\frac{5\sqrt{2}}{180}\frac{k}{2m_q}AP_3'(z)x^2$
&$+i\frac{5\sqrt{2}}{180}AP_2''(z)x^2$&
$-i\frac{5\sqrt{2}}{180}AP_3''(z)x^2$\\
\hline
$N(?)F_{15}$&$[70,^28,2,2]$&$+i\frac{2\sqrt{2}}{180}(1+\frac{k}{2m_q})AP_2'(z)x^2$
&$+i\frac{2\sqrt{2}}{180}\frac{k}{2m_q}AP_3'(z)x^2$
&$+i\frac{2\sqrt{2}}{180}AP_2''(z)x^2$&
$-i\frac{2\sqrt{2}}{180}AP_3''(z)x^2$\\
\hline
$\Delta(1950)F_{37}$&$[56,^410,2,2]$&$+i\frac{2\sqrt{2}}{3}\frac{k}{70m_q}AP_4'(z)x^2$
&$+i\frac{2\sqrt{2}}{3}\frac{2k}{105m_q}AP_3'(z)x^2$
&$-i\frac{2\sqrt{2}}{3}\frac{k}{210m_q}AP_4''(z)x^2$
&$+i\frac{2\sqrt{2}}{3}\frac{k}{210m_q}AP_3''(z)x^2$\\
\hline
\end{tabular}
\end{table}
\end{center}
\end{widetext}

\begin{widetext}
\begin{center}
\begin{table}[ht]
\caption{ The CGLN amplitudes of $s$-channel resonances in the
$n\leq2$ shell for the $\gamma n\rightarrow \pi^0 n$ process in the
SU(6)$\otimes$O(3) symmetry limit. } \label{CGLN2}
\begin{tabular}{|c|c|c|c|c|c|c|c|c|c| }\hline\hline
resonance &$[N_6,^{2S+1}N_3,n,l]$& $f^R_1$ \ \ & $f^R_2$ & $f^R_3$ & $f^R_4$  \\
\hline $N(940)P_{11}$& $[56,^28,0,0]$&0& $+i\frac{5\sqrt{2}}{3}\frac{k}{6m_q}A$ &0 &0\\
\hline $\Delta(1232)P_{33}$& $[56,^410,0,0]$&
$+i\frac{4\sqrt{2}}{3}\frac{k}{6m_q}A P_2'(z)$ &
$+i\frac{8\sqrt{2}}{3}\frac{k}{6m_q}A$
&$-i\frac{4\sqrt{2}}{3}\frac{k}{6m_q}A
P_2''(z)$ &  0\\
       \hline
$N(1535)S_{11}$&$[70,^28,1,1]$&
$-i\frac{\sqrt{2}}{18}k(1+\frac{k}{6m_q})B$
&0 &0 &  0\\
\hline $N(1650)S_{11}$&$[70,^48,1,1]$&
$+i\frac{\sqrt{2}}{36}k\frac{k}{6m_q}B$
&0 &0 &  0\\
\hline $\Delta(1620)S_{31}$&$[70,^210,1,1]$&
$+i\frac{\sqrt{2}}{36}k(1-\frac{k}{6m_q})B$
&0 &0 &  0\\
\hline $N(1520)D_{13}$&$[70,^28,1,1]$&
$+i\frac{\sqrt{2}}{9}(1+\frac{k}{6m_q})Ax$
&$+i\frac{\sqrt{2}}{9}\frac{k}{6m_q}AxP_2'(z)$ &0 &
$-i\frac{\sqrt{2}}{9}AxP_2''(z)$\\
\hline $N(1700)D_{13}$&$[70,^48,1,1]$&
$+i\frac{\sqrt{2}}{18}\frac{4}{5}\frac{k}{6m_q}Ax$
&$+i\frac{\sqrt{2}}{18}\frac{1}{5}\frac{k}{6m_q}AxP_2'(z)$ &0 &
$-i\frac{\sqrt{2}}{18}\frac{3}{5}\frac{k}{6m_q}AxP_2''(z)$\\
\hline
 $\Delta(1700)D_{33}$&$[70,^210,1,1]$&
$-i\frac{\sqrt{2}}{18}(1-\frac{k}{6m_q})Ax$
&$+i\frac{\sqrt{2}}{18}\frac{k}{6m_q}AxP_2'(z)$ &0 &
$+i\frac{\sqrt{2}}{18}AxP_2''(z)$\\
\hline $N(1675)D_{15}$&$[70,^48,1,1]$&
$+i\frac{\sqrt{2}}{6}\frac{k}{15m_q}AxP_3'(z)$
&$+i\frac{\sqrt{2}}{6}\frac{k}{10m_q}AxP_2'(z)$
&$-i\frac{\sqrt{2}}{6}\frac{k}{2m_q}Ax z$ &
$+i\frac{\sqrt{2}}{6}\frac{k}{30m_q}AxP_2''(z)$\\
\hline
 $N(1440)P_{11}$&$[56,^28,2,0]$& 0
&$+i\frac{47\sqrt{2}}{36\times108}\frac{10}{11}k\frac{k}{m_q}Cx$ &0
&
0\\
\hline $N(1710)P_{11}$&$[70,^28,2,0]$& 0
&$+i\frac{47\sqrt{2}}{36\times108}\frac{2}{11}k\frac{k}{m_q}Cx$ &0 &
0\\
\hline $\Delta(1750)P_{31}$&$[70,^210,2,0]$& 0
&$-i\frac{47\sqrt{2}}{36\times108}\frac{1}{11}k\frac{k}{m_q}Cx$ &0 &
0\\
\hline $N(?)P_{11}$&$[70,^48,2,2]$& 0
&$-i\frac{47\sqrt{2}}{36\times108}\frac{1}{9}k\frac{k}{m_q}Dx$ &0 &
0\\
\hline $\Delta(1910)P_{31}$&$[56,^410,2,2]$& 0
&$-i\frac{47\sqrt{2}}{36\times108}\frac{8}{9}k\frac{k}{m_q}Dx$ &0 &
0\\
\hline
$N(1720)P_{13}$&$[56,^28,2,2]$&$-i\frac{\sqrt{2}}{108}\frac{10}{2}k\frac{k}{6m_q}DP_2'(z)x$
&$-i\frac{\sqrt{2}}{108}\frac{10}{2}k\frac{k}{6m_q}Dx$ &0&
0\\
\hline
$N(1900)P_{13}$&$[70,^28,2,2]$&$-i\frac{\sqrt{2}}{108}k(1+\frac{k}{6m_q})DP_2'(z)x$
&$-i\frac{\sqrt{2}}{108}k\frac{k}{6m_q}Dx$
&$-i\frac{\sqrt{2}}{108}kDP_2''(z)x$&
0\\
\hline $N(?)P_{13}$&$[70,^48,2,2]$&0
&$-i\frac{\sqrt{2}}{108}\frac{1}{2}k\frac{k}{6m_q}Dx$
&$+i\frac{\sqrt{2}}{108}\frac{1}{2}k\frac{k}{6m_q}DP_2''(z)x$&
0\\
\hline
$\Delta(1985?)P_{33}$&$[70,^210,2,2]$&$+i\frac{\sqrt{2}}{108}\frac{1}{2}k(1-\frac{k}{6m_q})DP_2'(z)x$
&$-i\frac{\sqrt{2}}{108}\frac{1}{2}k\frac{k}{6m_q}Dx$
&$+i\frac{\sqrt{2}}{108}\frac{1}{2}kDP_2''(z)x$&
0\\
\hline $\Delta(1920)P_{33}$&$[56,^410,2,2]$&0
&$-i\frac{\sqrt{2}}{108}\frac{8}{2}k\frac{k}{6m_q}Dx$
&$+i\frac{\sqrt{2}}{108}\frac{8}{2}k\frac{k}{6m_q}DP_2''(z)x$&
0\\
\hline
$\Delta(1600)P_{33}$&$[56,^410,2,0]$&$+i\frac{\sqrt{2}}{108}\frac{8}{2}k\frac{k}{6m_q}CP_2'(z)x$
&$+i\frac{\sqrt{2}}{108}\frac{24}{2}k\frac{k}{6m_q}Cx$
&$-i\frac{\sqrt{2}}{108}\frac{16}{2}k\frac{k}{6m_q}CP_2''(z)x$&
0\\
\hline
$N(?)P_{13}$&$[70,^48,2,0]$&$+i\frac{\sqrt{2}}{108}\frac{1}{2}k\frac{k}{6m_q}CP_2'(z)x$
&$+i\frac{\sqrt{2}}{108}\frac{3}{2}k\frac{k}{6m_q}Cx$
&$-i\frac{\sqrt{2}}{108}\frac{2}{2}k\frac{k}{6m_q}CP_2''(z)x$&
0\\
\hline
$N(1680)F_{15}$&$[56,^28,2,2]$&$+i\frac{\sqrt{2}}{18}\frac{k}{6m_q}AP_2'(z)x^2$
&$+i\frac{\sqrt{2}}{18}\frac{k}{6m_q}AP_3'(z)x^2$ &0&
0\\
\hline
$N(?)F_{15}$&$[70,^28,2,2]$&$+i\frac{\sqrt{2}}{18}\frac{1}{5}(1+\frac{k}{6m_q})AP_2'(z)x^2$
&$+i\frac{\sqrt{2}}{18}\frac{1}{5}\frac{k}{6m_q}AP_3'(z)x^2$
&$+i\frac{\sqrt{2}}{18}\frac{1}{5}AP_2''(z)x^2$&
$-i\frac{\sqrt{2}}{18}\frac{1}{5}AP_3''(z)x^2$\\
\hline
$N(?)F_{15}$&$[70,^48,2,2]$&$+i\frac{\sqrt{2}}{18}\frac{1}{14}\frac{k}{6m_q}AP_2'(z)x^2$
&$+i\frac{\sqrt{2}}{18}\frac{1}{35}\frac{k}{6m_q}AP_3'(z)x^2$
&$+i\frac{\sqrt{2}}{18}\frac{3}{70}\frac{k}{6m_q}AP_2''(z)x^2$&
$-i\frac{\sqrt{2}}{18}\frac{3}{70}\frac{k}{6m_q}AP_3''(z)x^2$\\
\hline
$\Delta(?)F_{35}$&$[70,^210,2,2]$&$-i\frac{\sqrt{2}}{18}\frac{1}{10}(1-\frac{k}{6m_q})AP_2'(z)x^2$
&$+i\frac{\sqrt{2}}{18}\frac{1}{10}\frac{k}{6m_q}AP_3'(z)x^2$
&$-i\frac{\sqrt{2}}{18}\frac{1}{10}AP_2''(z)x^2$&
$+i\frac{\sqrt{2}}{18}\frac{1}{10}AP_3''(z)x^2$\\
\hline
$\Delta(1905)F_{35}$&$[56,^410,2,2]$&$i\frac{\sqrt{2}}{18}\frac{4}{7}\frac{k}{6m_q}AP_2'(z)x^2$
&$+i\frac{\sqrt{2}}{18}\frac{8}{35}\frac{k}{6m_q}AP_3'(z)x^2$
&$+i\frac{\sqrt{2}}{18}\frac{12}{35}\frac{k}{6m_q}AP_2''(z)x^2$&
$-i\frac{\sqrt{2}}{18}\frac{12}{35}\frac{k}{6m_q}AP_3''(z)x^2$\\
\hline
$\Delta(1950)F_{37}$&$[56,^410,2,2]$&$+i\frac{3\sqrt{2}}{4}\frac{8}{9}\frac{k}{70m_q}AP_4'(z)x^2$
&$+i\frac{3\sqrt{2}}{4}\frac{8}{9}\frac{2k}{105m_q}AP_3'(z)x^2$
&$-i\frac{3\sqrt{2}}{4}\frac{8}{9}\frac{k}{210m_q}AP_4''(z)x^2$&
$+i\frac{3\sqrt{2}}{4}\frac{8}{9}\frac{k}{210m_q}AP_3''(z)x^2$\\
\hline
$N(1990)F_{17}$&$[70,^48,2,2]$&$+i\frac{3\sqrt{2}}{4}\frac{1}{9}\frac{k}{70m_q}AP_4'(z)x^2$
&$+i\frac{3\sqrt{2}}{4}\frac{1}{9}\frac{2k}{105m_q}AP_3'(z)x^2$
&$-i\frac{3\sqrt{2}}{4}\frac{1}{9}\frac{k}{210m_q}AP_4''(z)x^2$&
$+i\frac{3\sqrt{2}}{4}\frac{1}{9}\frac{k}{210m_q}AP_3''(z)x^2$\\
\hline
\end{tabular}
\end{table}
\end{center}
\end{widetext}

Finally, the differential cross section $d\sigma/d\Omega$, photon
beam asymmetry $\Sigma$, polarization of recoil protons $P$, and
target asymmetry $T$ are given by the following standard
expressions~\cite{Walker:1968xu,Krusche:2003ik,Fasano:1992es}:
\begin{eqnarray}
\frac{d\sigma}{d\Omega}&=&\frac{\alpha_e\alpha_{\pi}(E_i+M_N)(E_f+M_N)}{16s
M_N^2}\frac{1}{2}\frac{|\mathbf{q}|}{|\mathbf{k}|}\sum^4_{i=1}|H_i|^2,\label{dfcsc}\\
\Sigma &=&2\mathrm{Re}(H_4^*H_1-H_3^*H_2)/\sum^4_{i=1}|H_i|^2,\\
P &=&-2\mathrm{Im}(H_4^*H_2+H_3^*H_1)/\sum^4_{i=1}|H_i|^2,\\
T &=&2\mathrm{Im}(H_2^*H_1+H_4^*H_3)/\sum^4_{i=1}|H_i|^2,
\end{eqnarray}
where the transition amplitudes $H_i$ in the helicity space can be
expressed by the CGLN amplitudes $f_i$~\cite{Walker:1968xu}:
\begin{eqnarray}
H_1&=&-\frac{1}{\sqrt{2}}\sin\theta\cos\frac{\theta}{2}(f_3+f_4),\\
H_2&=&\sqrt{2}\cos\frac{\theta}{2}[(f_2-f_1)+\sin^2\frac{\theta}{2}(f_3-f_4)],\\
H_3&=&\frac{1}{\sqrt{2}}\sin\theta\sin\frac{\theta}{2}(f_3-f_4),\\
H_4&=&\sqrt{2}\sin\frac{\theta}{2}[(f_2+f_1)+\cos^2\frac{\theta}{2}(f_3-f_4)].
\end{eqnarray}
In Eq.(\ref{dfcsc}), the fine-structure constant $\alpha_e$ is well
determined, and the $\pi NN$ coupling constant $\alpha_{\pi}$ is
related to the axial vector coupling $g_A$ by the generalized
Goldberg-Treiman relation
\begin{eqnarray}
\alpha_{\pi}=\frac{1}{4\pi}\left(\frac{g_AM_N}{f_\pi}\right)^2\equiv
\frac{g_{\pi NN}^2}{4\pi}.
\end{eqnarray}
However, the quark model predicts rather large values $g_A=5/3$ for
charged pions and $g_A=5\sqrt{2}/6$ for neutral pions. In our paper,
the coupling $\alpha_{\pi}$ is determined by fitting the data.

\section{CALCULATION AND ANALYSIS}\label{discuss}

\subsection{Parameters}

In our framework, the $s$-channel resonance transition amplitude,
$\mathcal{O}_R$, is derived in the SU(6)$\otimes$O(3) symmetry
limit. In reality, the SU(6)$\otimes$O(3) symmetry is generally
broken due to, e.g., spin-dependent forces in the quark-quark
interaction. As a consequence, configuration mixings would occur.
The configuration mixings break the SU(6)$\otimes$O(3) symmetry,
which can change our theoretical predictions. Furthermore, the
helicity couplings and strong decay couplings of some resonances
might be over/under-estimated with the simple quark model. To
accommodate the uncertainties in the symmetric quark model
framework, we introduce a set of coupling strength parameters,
$C_R$, for each resonance amplitude by an empirical
method~\cite{Li:1998ni, Saghai:2001yd,He:2008ty,He:2009zzi}:
\begin{eqnarray}
\mathcal{O}_R\rightarrow C_{R} \mathcal{O}_{R},
\end{eqnarray}
where $C_{R}$ can be determined by fitting the experimental
observables. In the SU(6)$\otimes$O(3) symmetry limit one finds
$C_{R}=1$, while deviations of $C_{R}$ from unity imply the
SU(6)$\otimes$O(3) symmetry breaking.

In our previous study of the $\eta$ photoproduction on the nucleons,
we found the configuration mixings seem to be inevitable for the
low-lying $S$-wave nucleon resonances $N(1535)S_{11}$ and
$N(1650)S_{11}$, and $D$-wave nucleon resonances $N(1520)D_{13}$ and
$N(1700)D_{13}$. By including configuration mixing effects in the
$S$- and $D$-wave states, we explicitly express their transition
amplitudes as follows:
\begin{eqnarray}\label{coeff-1}
\mathcal{O}_R\rightarrow C^{[70,^{2}8]}_{R}
\mathcal{O}_{[70,^{2}8,J]}+C^{[70,^{4}8]}_{R}
\mathcal{O}_{[70,^{4}8,J]}.
\end{eqnarray}
The coefficients $C^{[70,^{2}8]}_{R}$ and $C^{[70,^{4}8]}_{R}$ can
be related to the mixing angles.
We adopt the same mixing scheme as in our previous
work~\cite{Zhong:2011ti},
\begin{equation}\label{mixs}
\left(\begin{array}{c}S_{11}(1535)\cr S_{11}(1650)
\end{array}\right)=\left(\begin{array}{cc} \cos\theta_{S} &-\sin\theta_{S}\cr \sin\theta_{S} & \cos\theta_{S}
\end{array}\right)
\left(\begin{array}{c} \left|70,^28, 1/2^-\right\rangle \cr
\left|70,^48, 1/2^-\right\rangle
\end{array}\right),
\end{equation}
and
\begin{equation}\label{mixd}
\left(\begin{array}{c}D_{13}(1520)\cr D_{13}(1700)
\end{array}\right)=\left(\begin{array}{cc} \cos\theta_{D} &-\sin\theta_{D}\cr \sin\theta_{D} & \cos\theta_{D}
\end{array}\right)
\left(\begin{array}{c} \left|70,^28, 3/2^-\right\rangle \cr
\left|70,^48, 3/2^-\right\rangle
\end{array}\right).
\end{equation}
Then, the coefficients defined in Eq.(\ref{coeff-1}) are given by
\begin{eqnarray}\label{mix-coeff1}
C^{[70,^28]}_{S_{11}(1535)}&=&R_2^S\cos\theta_S(\cos\theta_S-\sin\theta_S/2),\label{mix-c1}\\
C^{[70,^48]}_{S_{11}(1535)}&=&R_4^S\sin\theta_S(\sin\theta_S-2\cos\theta_S),\\
C^{[70,^28]}_{S_{11}(1650)}&=&R_2^S\sin\theta_S(\sin\theta_S+\cos\theta_S/2),\\
C^{[70,^48]}_{S_{11}(1650)}&=&R_4^S\cos\theta_S(\cos\theta_S+2\sin\theta_S),\\
C^{[70,^28]}_{D_{13}(1520)}&=&R_2^D\cos\theta_D(\cos\theta_D-\frac{1}{2\sqrt{10}}\sin\theta_D),\\
C^{[70,^48]}_{D_{13}(1520)}&=&R_4^D\sin\theta_D(\sin\theta_D-2\sqrt{10}\cos\theta_D),\\
C^{[70,^28]}_{D_{13}(1700)}&=&R_2^D\sin\theta_D(\sin\theta_D+\frac{1}{2\sqrt{10}}\cos\theta_D),\\
C^{[70,^48]}_{D_{13}(1700)}&=&R_4^D\cos\theta_D(\cos\theta_D+2\sqrt{10}\sin\theta_D)\label{mix-c2}.
\end{eqnarray}
The parameters $R_2$ and $R_4$ are introduced to adjust the overall
strength of the partial wave amplitudes of $[70,^28]$ and
$[70,^48]$, respectively, which may be overestimated or
underestimated in the naive quark model~\cite{Zhong:2011ti}. If
$R=1$, one finds that the $C_R$ parameters of $S$- and $D$-wave
states can be explained with configuration mixings only. In the
calculation, the mixing angle between $N(1535)S_{11}$ and
$N(1650)S_{11}$ is adopted to be $\theta_S=26^\circ$, determined in
our previous work~\cite{Zhong:2011ti}. Notice that in our work, we
have adopted Isgur's later conventions~\cite{Koniuk:1979vy} where
wave functions are in line with the SU(3) conventions of de
Swart~\cite{deSwart:1963gc}. In this frame, we obtain a positive
mixing angle $\theta_S$. However, in line with the old conventions
of the SU(3) wave functions from Isgur and Karl's early
works~\cite{Isgur:1978xj,Isgur:1977ef}, one obtains a negative
mixing angle
$\theta_S$~\cite{Li:1998ni,Saghai:2001yd,He:2008ty,He:2009zzi,Isgur:1978xj,Hey:1974nc}.
This question has been clarified in
Refs.~\cite{An:2011sb,Zhong:2011ti}. Furthermore, the mixing angle
between $N(1520)D_{13}$ and $N(1700)D_{13}$ is adopted to be
$\theta_D\simeq 10^\circ$ as widely suggested in the
literature~\cite{Hey:1974nc,Capstick:2004tb,Isgur:1978xj,Saghai:2001yd,He:2008ty,He:2009zzi}.


For the $\gamma p\to \pi^0 p$ reaction, we obtain $R_2^S\simeq 1.0$
and $R_2^D\simeq 1.5$ by fitting the 450 data points of differential
cross section, and the 53 data points of total cross section
collected in Tab.~\ref{xsq}. While for the $\gamma n\to \pi^0 n$
reaction, we obtain $R_2^S\simeq R_4^S\simeq 0.7$, $R_2^D\simeq 1.4$
and $R_4^D\simeq 1.0$ by fitting the 36 data points of total cross
section around the second resonance energy region $1.30\leq W \leq
1.72$ GeV recently measured at MAMI~\cite{Dieterle:2014blj}. With
these determined $R$ parameters, from Eqs.~\ref{mix-c1}-\ref{mix-c2}
one can obtain the overall strength parameters $C^{[70,^{2}8]}_{R}$
and $C^{[70,^{4}8]}_{R}$ for the $S$- and $D$-wave resonances.

The determined $C_R$ values for these low-lying resonances are
listed in Tab.~\ref{param C}. From the table, we find that to
reproduce the data we need to introduce two large coupling strength
parameters $C_{P_{33}(1232)}\simeq 1.83$ and $C_{S_{31}(1620)}\simeq
1.8$ for $\Delta(1232)P_{33}$ and $\Delta(1620)S_{31}$,
respectively. The reason may be the well-known underestimation of
their photocouplings in the constituent quark
model~\cite{Sato:1996gk,Sato:2000jf}. We also need to enhance the
contributions of $N(1520)D_{13}$ by a factor of
$C_{D_{13}(1520)}\simeq 1.4$, which can not be explained with
configuration mixings only. The underestimation of the resonance
amplitude of $N(1520)D_{13}$ is also found in the $\gamma N\to \eta
N$ processes within the quark model framework~\cite{Zhong:2011ti},
which is due to the underestimation of the photocoupling of
$N(1520)D_{13}$ in the constituent quark model. In the $\pi^0$
photoproduction processes, the data favor a smaller contribution of
$N(1535)S_{11}$ than that in the SU(6)$\otimes$O(3) symmetry limit.
In the $\gamma p\to \pi^0 p$ reaction, the strength parameter
$C^{[70,^28]}_{S_{11}(1535)}\simeq 0.61$ can be naturally explained
with the configuration mixings between two $S$-wave nucleon
resonances $N(1535)S_{11}$ and $N(1650)S_{11}$ with a mixing angle
$\theta_S\simeq26^\circ$. However, in the $\gamma n\to \pi^0 n$
reaction, the strength parameter $C^{[70,^28]}_{S_{11}(1535)}\simeq
0.43$ can not be well explained with the configuration mixing
effects only, we should introduce a parameter $R_2^S\simeq 0.7$ to
slightly decrease the transition amplitude of $[70,^28,1/2^-]$.
Furthermore, we find that in the $\gamma n\to \pi^0 n$ reaction, the
enhancement of the contributions of $N(1720)P_{13}$ might
significantly improve the descriptions of the experimental data.

\begin{table}[ht]
\caption{450 data points of differential cross section, and 53 data
points of total cross section of $\gamma p\rightarrow \pi^0 p$
included in our fits. The $\chi^2$ datum point is about
$\chi^2/N_{data}=4.3$. } \label{xsq}
\begin{tabular}{|c|c|c|c|c|c| }\hline\hline
 Data Refs. &  Obser.  & $E_\gamma$ &  $N_{data}$ & $\chi^2_i$ & $\chi^2_i/N_{data}$ \\
\hline
\cite{Fuchs:1996ja} MAMI\ \ \ \ \ \ & $d\sigma/d\Omega$   & $240,260,278$      &27 & 357 & 13.2\\
\cite{Beck:2006ye}  MAMI\ \ \ \ \ \ &  $d\sigma/d\Omega$ & $300\sim 400$ & 112 & 540 & 4.8\\
\cite{Bartholomy:2004uz} CB-ELSA & $d\sigma/d\Omega$ & $438\sim 862$  & 311 & 1204 & 3.9\\
\cite{Schumann:2010js}  MAMI\ \ \ \ \ \ & $\sigma$  & $240\sim 335$ & 20 & 10 & 0.5\\
\cite{Bartholomy:2004uz} CB-ELSA & $\sigma$  & $342\sim 1138$  & 33 & 37 & 1.1\\
\hline
\end{tabular}
\end{table}

\begin{table}[ht]
\caption{The strength parameters $C_R$ determined by the
experimental data.} \label{param C}
\begin{tabular}{|c|c|c|c|c|c|c }\hline\hline
$C_R$ parameter &\ \  $\gamma p\rightarrow \pi^0 p$ \ \ & \ \
$\gamma
n\rightarrow \pi^0 n$   \\
\hline
$C^{[70,^28]}_{S_{11}(1535)}$     & $0.61^{+0.06}_{-0.04}$     &$0.43^{+0.09}_{-0.09}$        \\
$C^{[70,^48]}_{S_{11}(1535)}$     & ...     &$-0.42$        \\
$C^{[70,^28]}_{S_{11}(1650)}$     & $0.39^{+0.04}_{-0.06}$    &$0.27^{+0.06}_{-0.06}$        \\
$C^{[70,^48]}_{S_{11}(1650)}$     & ...     &1.12        \\
$C^{[70,^28]}_{D_{13}(1520)}$     & $1.41^{+0.15}_{-0.09}$    &$1.32^{+0.09}_{-0.09}$        \\
$C^{[70,^48]}_{D_{13}(1520)}$     & ...     &$-1.05$       \\
$C^{[70,^28]}_{D_{13}(1700)}$     & 0.09    &0.08        \\
$C^{[70,^48]}_{D_{13}(1700)}$     & ...     &2.05        \\
$C_{P_{33}(1232)}$     & $1.83^{+0.02}_{-0.04}$     &$1.83 $      \\
$C_{S_{31}(1620)}$     & $1.80^{+0.50}_{-0.20}$     &$1.80 $         \\
$C_{D_{15}(1675)}$     & 1.00     &1.00         \\
$C_{P_{13}(1720)}$     & 1.00     &$3.20^{+0.1}_{-0.2}$         \\
\hline
\end{tabular}
\end{table}

To take into account relativistic effects, the commonly applied
Lorentz boost factor is introduced in the resonance amplitude for
the spatial integrals~\cite{Li:1995si}, which is
\begin{eqnarray}
\mathcal{O}_R(\textbf{k},\textbf{q})\rightarrow
\mathcal{O}_R(\gamma_k\textbf{k}, \gamma_q\textbf{q} ),
\end{eqnarray}
where $\gamma_k=M_N/E_i$ and $\gamma_q=M_N/E_f$.

The $\pi NN$ coupling $\alpha_\pi$ and the coupling $g_{\omega \pi
\gamma}\cdot g_{\omega qq}$ from $\omega$-meson exchange in the $t$
channel are considered as free parameters in the present
calculations. By fitting the experimental data of $\gamma
p\rightarrow \pi^0 p$ reaction (see Tab.~\ref{xsq}), we get $g_{\pi
NN}\simeq 13.2$ (i.e., $\alpha_\pi\equiv g^2_{\pi
NN}/4\pi\simeq13.8$) and $g_{\omega\pi \gamma}\cdot g_{\omega
qq}\simeq 1.37$. The $\pi NN$ coupling determined in this work is
compatible with the value $g_{\pi NN}\simeq 13.5$ adopted in other
literature~\cite{Huang:2011as,Ronchen:2014cna}. According to the
decay of $\omega\to \pi \gamma$, one obtains $g_{\omega \pi
\gamma}\simeq 0.32$~\cite{Zhao:2002id}. Then the $\omega qq$
coupling extracted by us is $g_{\omega qq}\simeq 4.28$, which is
consistent with the value $g_{\omega qq}\simeq 3$ suggested in
Ref.~\cite{Riska:2000gd}.

There are another two parameters, the constituent quark mass $m_q$
and the harmonic oscillator strength $\alpha$, from the transition
amplitudes. In the calculation we adopt their standard values in the
the quark model, $m_q=330$ MeV and $\alpha^2=0.16$ GeV$^2$.

In the calculations, the $n=3$ shell resonances are treated as
degeneration; their degenerate mass and width are taken as $M=2080$
MeV and $\Gamma=200$ MeV, since in the low energy region the
contributions from the $n=3$ shell are not significant. In the $u$
channel, the intermediate states are the nucleon and $\Delta(1232)$
and their resonances. It is found that contributions from the $n\geq
1$ shell are negligibly small and insensitive to the degenerate
masses for these shells. In this work, we take $M_1=1650$ MeV
($M_2=1750$ MeV) for the degenerate mass of $n=1$ ($n=2$) shell
resonances. In the $s$ channel, the masses and widths of the
resonances are taken from the PDG~\cite{Agashe:2014kda}, or the
constituent quark model predictions~\cite{Capstick:1986bm} if no
experimental data are available. For the main resonances, we allow
their masses and widths to change around the values from
PDG~\cite{Agashe:2014kda} in order to better describe the data. The
determined values are listed in Tab. \ref{parameter}. As a
comparison, the resonance masses and widths of both pole and
Breit-Wigner parametrizations from the PDG~\cite{Agashe:2014kda} are
listed in Tab. \ref{parameter} as well. It is found that the
resonance masses and widths extracted by us are in good agreement
with the values of pole parametrization. The reason is that when we
fit the data, a momentum independent width $\Gamma_R$ is used, which
is similar to the pole parametrization. It should be pointed out
that $N(1720)P_{13}$ seems to be a narrow state with a width of $120
$ MeV in our model, which is about one half of the average value
from the PDG~\cite{Agashe:2014kda}. However, our result is in good
agreement with that extracted from the $\pi^-p\to K^0 \Lambda$
reaction by D. H. Saxon {\it et al.}~\cite{Saxon:1979xu}. The strong
decay properties of $N(1720)P_{13}$ will be discussed in detail in
our another work.

To know some uncertainties of a main parameter ($C_R$, $M_R$,
$\Gamma_R$) we vary it around its central value until the
predictions are inconsistent with the data within their
uncertainties. The obtained uncertainties for the main parameters
have been given in Tabs.~\ref{param C} and ~\ref{parameter}.

\begin{table}[ht]
\caption{The masses $M_R$ (MeV) and widths $\Gamma_R$ (MeV) for the
$s$-channel resonances in present work compared with the values of
Breit-Wigner (BW) and pole parametrizations from
PDG14~\cite{Agashe:2014kda}. } \label{parameter}
\begin{tabular}{|c|c|c|c|c|c| }\hline\hline
resonance &  $(M_R,\Gamma_R)_{\mathrm{Ours}}$  &  $(M_R,\Gamma_R)_{\mathrm{Pole}}$ &$(M_R,\Gamma_R)_{\mathrm{BW}}$   \\
\hline
$\Delta(1232)P_{33}$& $1210^{+2}_{-2},98^{+2}_{-2}$   &$1210^{+1}_{-1},100^{+2}_{-2}$       &$1232^{+2}_{-2},117^{+3}_{-3}$   \\
$N(1535)S_{11}$     & $1515^{+5}_{-7},115^{+10}_{-15}$  &$1510^{+20}_{-20},170^{+80}_{-80} $  &$1535^{+10}_{-10},150^{+25}_{-25}$   \\
$N(1650)S_{11}$     & $1660^{+10}_{-20},150^{+30}_{-20}$  &$1655^{+15}_{-15}, 135^{+35}_{-35}$  &$1655^{+15}_{-10},140^{+30}_{-30} $    \\
$\Delta(1630)S_{31}$& $1600^{+10}_{-10},135^{+25}_{-15}$  &$1600^{+10}_{-10},130^{+10}_{-10}$   &$1630^{+30}_{-30},140^{+10}_{-10} $      \\
$N(1520)D_{13}$     & $1518^{+5}_{-5},105^{+5}_{-10}$  &$1510^{+5}_{-5},110^{+10}_{-5}$      &$1515^{+5}_{-5},115^{+10}_{-15}$   \\
$N(1720)P_{13}$     & $1685^{+10}_{-5},120^{+5}_{-10}$  &$1675^{+15}_{-15},250^{+150}_{-100}$ &$1720^{+30}_{-20}, 250^{+150}_{-100}$   \\
\hline
\end{tabular}
\end{table}

\subsection{$\gamma p\rightarrow \pi^0 p$}

The chiral quark model studies of $\gamma p\rightarrow \pi^0 p$ were
carried out in Refs.~\cite{Li:1994cy,Li:1997gd,Zhao:2002id} about
twenty years ago. During the past two decades, great progress has
been achieved for pion photoproduction at JLab, CB-ELSA, MAMI, and
GRAAL. The new data sets are more accuracy and have larger solid
angle coverage and wider photon energy range. The improvement of the
experimental situations gives us a good opportunity to test our
model and study the excitation spectra of nucleon and $\Delta(1232)$
at the same time. All the intermediate states in the $s$ channel
classified in the quark model with $n\leq 2$ are listed in
Tab.~\ref{CGLN1}. It should be pointed out that in this reaction the
contributions from the nucleon excitations with the representation
$[70, ^48]$ are forbidden by the Moorhouse selection
rule~\cite{Moorhouse:1966jn,Zhao:2006an}. In the $n=0$ shell, both
nucleon pole and $\Delta(1232)P_{33}$ contribute to the reaction.
Comparing their CGLN amplitudes listed in Tab.~\ref{CGLN1}, we can
obviously see that $\Delta(1232)P_{33}$ plays a dominant role for
its larger amplitudes. In the $n=1$ shell, two $S$-wave states
$N(1535)S_{11}$ and $\Delta(1620)S_{31}$ and two $D$-wave states
$N(1520)D_{13}$ and $\Delta(1700)D_{33}$ contribute to the reaction.
Considering configuration mixing effects, we find that
$N(1650)S_{11}$ and $N(1700)D_{13}$ can also contribute to the
reaction. Similarly, from Tab.~\ref{CGLN1} we can find that
$N(1535)S_{11}$ and $N(1520)D_{13}$ play a dominant role in the
$n=1$ shell $S$-wave and $D$-wave resonances, respectively. In the
$n=2$ shell eight $P$-wave resonances and five $F$-wave resonances
contribute to the reaction. Comparing their CGLN amplitudes we find
that $N(1720)P_{13}$ and $N(1680)F_{15}$ play a dominant role in the
$n=2$ shell $P$-wave resonances and $F$-wave resonances,
respectively.

\begin{figure}[ht]
\centering \epsfxsize=8.6 cm \epsfbox{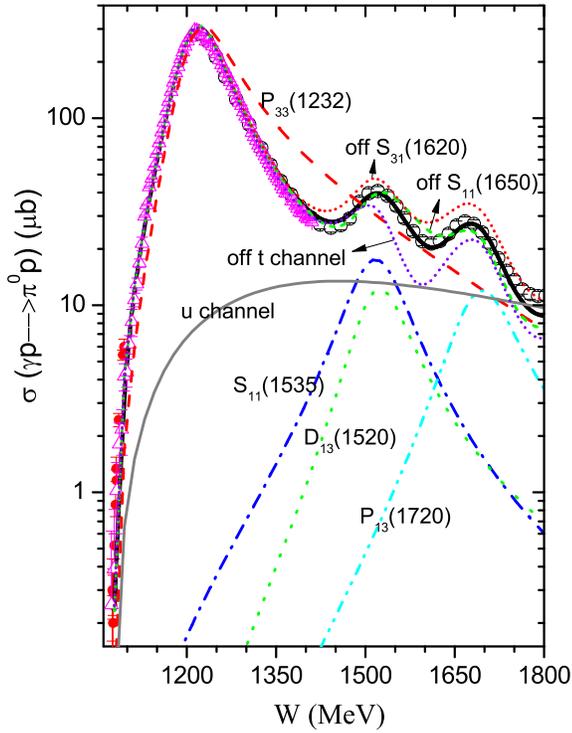} \caption{(Color
online) Total cross section as a function of c.m. energy $W$ for the
$\gamma p \rightarrow \pi^0p$ reaction. Data are taken from
\cite{Bartholomy:2004uz} (open circles), \cite{Mazzucato:1986dz}
(solid circles), and \cite{Schumann:2010js} (triangles). The results
for switching off the contributions from $N(1650)S_{11}$,
$\Delta(1620)S_{31}$, and $t$ channel; and the partial cross
sections for $\Delta(1232)P_{33}$, $N(1535)S_{11}$, $N(1520)D_{13}$,
$N(1720)P_{13}$ and the $u$ channel are indicated explicitly by
different legends in the figure. }\label{p3}
\end{figure}
\begin{figure}[ht]
\centering \epsfxsize=8.6 cm \epsfbox{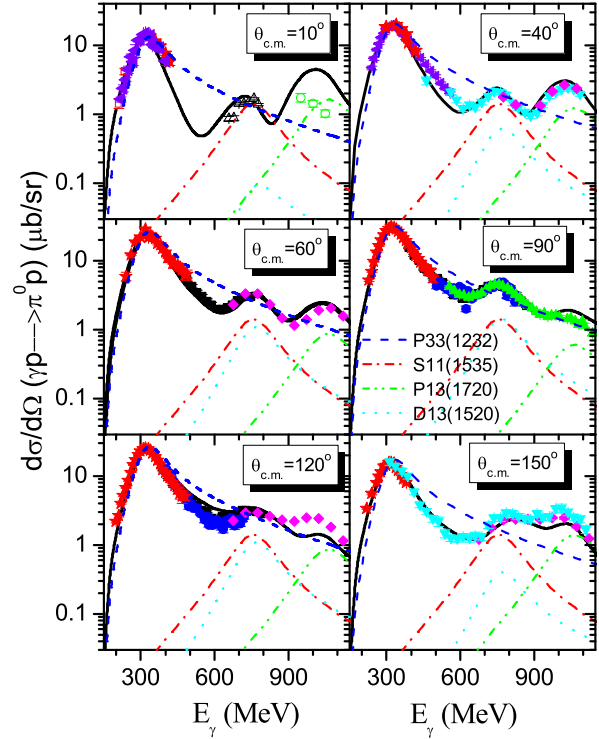} \caption{(Color
online) Photon energy dependent differential cross section of the
$\gamma p \rightarrow \pi^0p$ reaction. Data are taken
from~\cite{Beck:2006ye} (left triangles),~\cite{Govorkov:1966} (open
squares),~\cite{Fischer:1970dn} (solid
stars),~\cite{DeStaebler:1965sf} (open triangles),
\cite{Hemmi:1973nf} (open circles), \cite{Yoshioka:1980vu} (solid
squares),~\cite{Bacci:1967zz} (solid
circles),~\cite{Bartholomy:2004uz} (down triangles),
\cite{Bartalini:2005wx} (up triangles),~\cite{Dugger:2007bt}
(diamonds). The partial cross sections for $\Delta(1232)P_{33}$,
$N(1535)S_{11}$, $N(1520)D_{13}$ and $N(1720)P_{13}$ are indicated
explicitly by different legends in the figure.}\label{p2}
\end{figure}
\begin{figure}[ht]
\centering \epsfxsize=8.6 cm \epsfbox{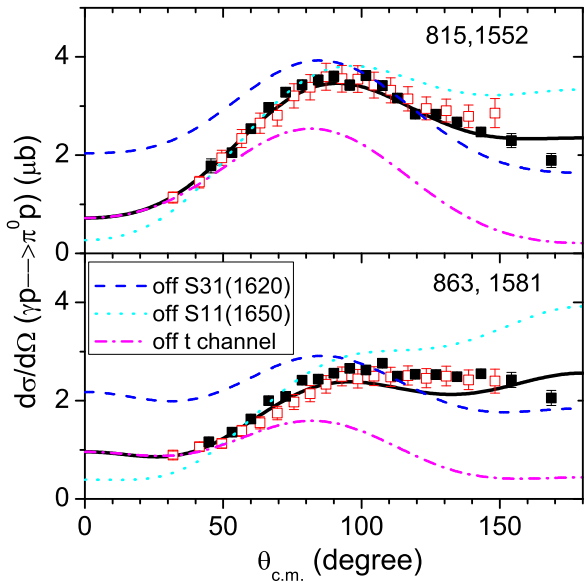} \caption{(Color
online) Effects of $N(1650)S_{11}$, $\Delta(1620)S_{31}$ and $t$
channel on the differential cross sections of the $\gamma
p\rightarrow \pi^0 p$ process. Data are taken
from~\cite{Bartholomy:2004uz} (open squares) and
\cite{Bartalini:2005wx} (solid squares). The results by switching
off the contributions from $N(1650)S_{11}$, $\Delta(1620)S_{31}$,
and $t$ channel are indicated explicitly by different legends in the
figure. The first and second number in each figure correspond to the
photon energy $E_\gamma$ (MeV) and the $\pi N$ center-of-mass energy
$W$ (MeV), respectively.}\label{peff}
\end{figure}
\begin{figure}[ht]
\centering \epsfxsize=8.6 cm \epsfbox{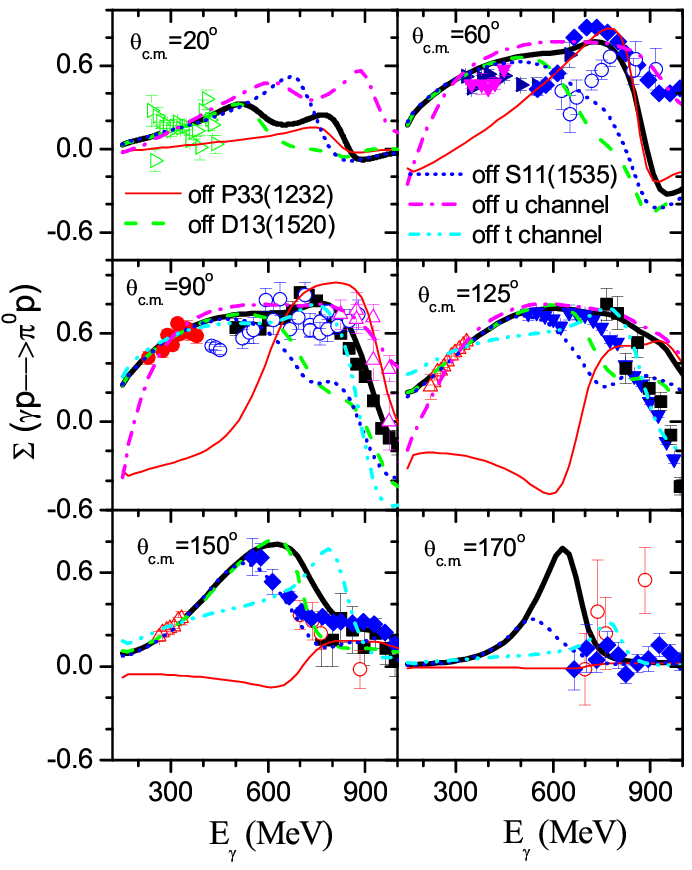} \caption{(Color
online) Photon energy dependent beam asymmetry of the $\gamma p
\rightarrow \pi^0p$ reaction. The data are taken from
\cite{Ganenko:1974zp} (right triangles), \cite{Blanpied:2001ae}
(open up triangles), \cite{Belyaev:1983xf} (open diamonds),
\cite{Bartalini:2005wx} (solid diamonds), \cite{Adamian:2000yi}
(down triangles), \cite{Knies:1974zx} (open circles),
\cite{Elsner:2008sn} (solid squares), \cite{Barbiellini:1970qu}
(solid circles), and \cite{Zdarko:1972di} (open squares). The
results by switching off the contributions from various partial
waves are indicated explicitly by different legends in the figure.
}\label{p5}
\end{figure}
\begin{figure}[ht]
\centering \epsfxsize=8.6 cm \epsfbox{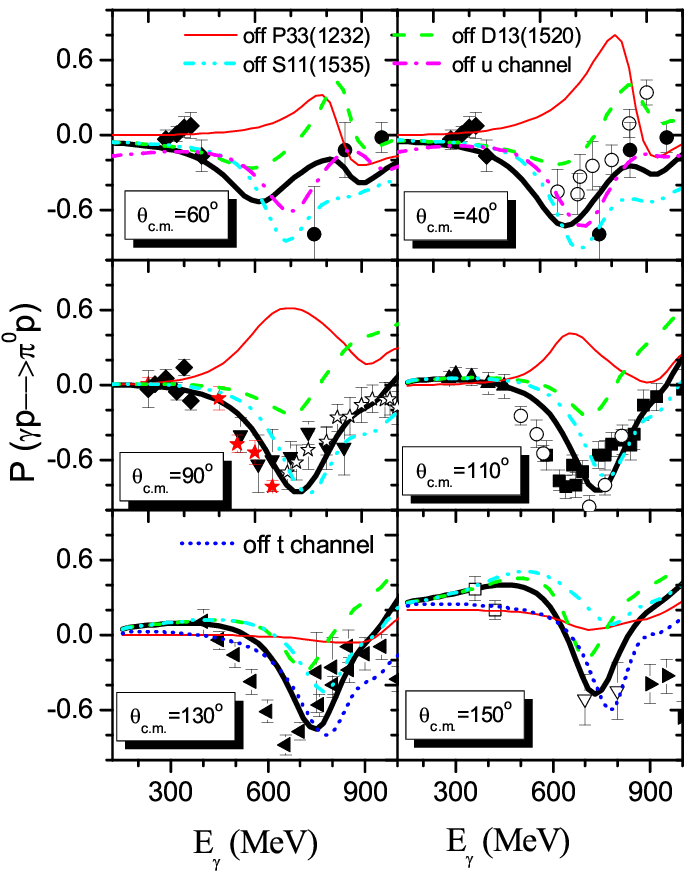} \caption{(Color
online) Photon energy dependent polarization of recoil protons for
the $\gamma p \rightarrow \pi^0p$ reaction. Data are taken from
\cite{Lundquist:1968zz} (open circles), \cite{Bloom:1967tn}(solid
circles), \cite{Belyaev:1982zv} (diamonds), \cite{Querzoli:1961zya}
(solid down triangles), \cite{Althoff:1966} (solid stars),
\cite{Goncharov:1973we} (open stars), \cite{Bratashevsky:1985pk}
(solid squares), \cite{Belyaev:1983xf} (solid up triangles),
\cite{Kato:1979br} (solid left triangles), \cite{Blum:1976mw} (open
down triangles), and \cite{Blum:1976ud} (solid right triangles). The
results by switching off the contributions from various partial
waves are indicated explicitly by different legends in the figure.
}\label{p7}
\end{figure}
\begin{figure}[ht]
\centering \epsfxsize=8.6 cm \epsfbox{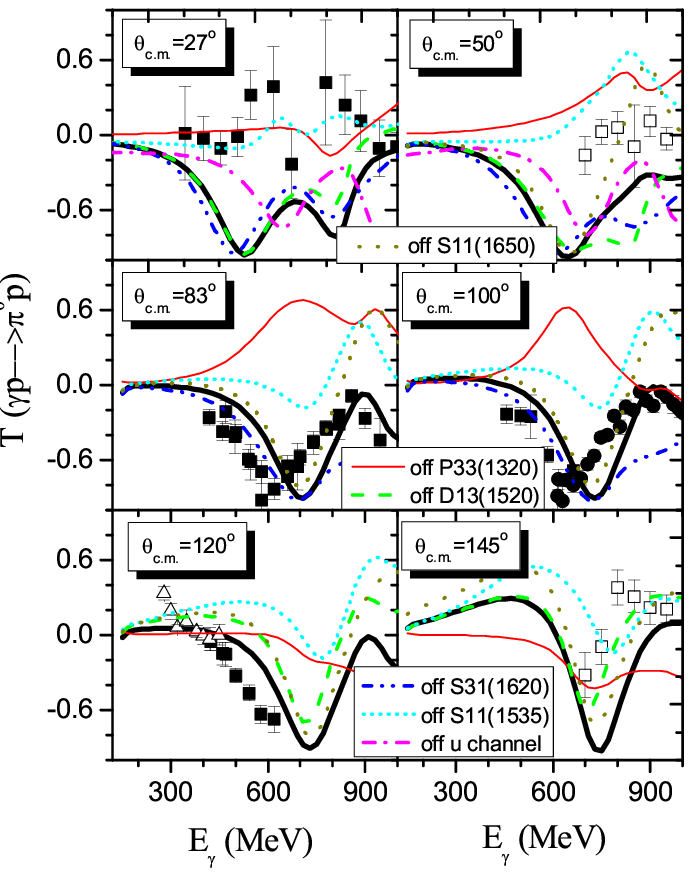} \caption{(Color
online) Photon energy dependent target asymmetry of the $\gamma p
\rightarrow \pi^0p$ reaction. The data are taken from
\cite{Fukushima:1977xj} (solid squares), \cite{Booth:1976es} (open
squares), \cite{Feller:1976ta} (solid circles), and
\cite{Belyaev:1983xf} (open triangles). The results by switching off
the contributions from various partial waves are indicated
explicitly by different legends in the figure.}\label{p9}
\end{figure}

In present work, we have calculated the differential cross sections,
total cross section, beam asymmetry, target asymmetry, and
polarization of recoil protons from pion production threshold up to
the second resonance region for the $\gamma p\rightarrow \pi^0 p$
reaction. The model parameters are determined by fitting the 450
data points of differential cross section from MAMI
\cite{Fuchs:1996ja,Beck:2006ye} and CB-ELSA \cite{Bartholomy:2004uz}
in the beam energy region 240 MeV$\leq E_\gamma \leq 862$ MeV, and
the 53 data points of total cross section from MAMI
\cite{Schumann:2010js} and CB-ELSA \cite{Bartholomy:2004uz} in the
beam energy region $240\leq E_\gamma \leq 1138$ MeV (see
Tab.~\ref{xsq}). The $\chi^2$ datum point is about
$\chi^2/N_{data}=4.3$. Our results are compared with the data in
Figs~\ref{p1}-\ref{p9}.

\begin{widetext}
\begin{center}
\begin{figure}[ht]
\centering \epsfxsize=17.2 cm \epsfbox{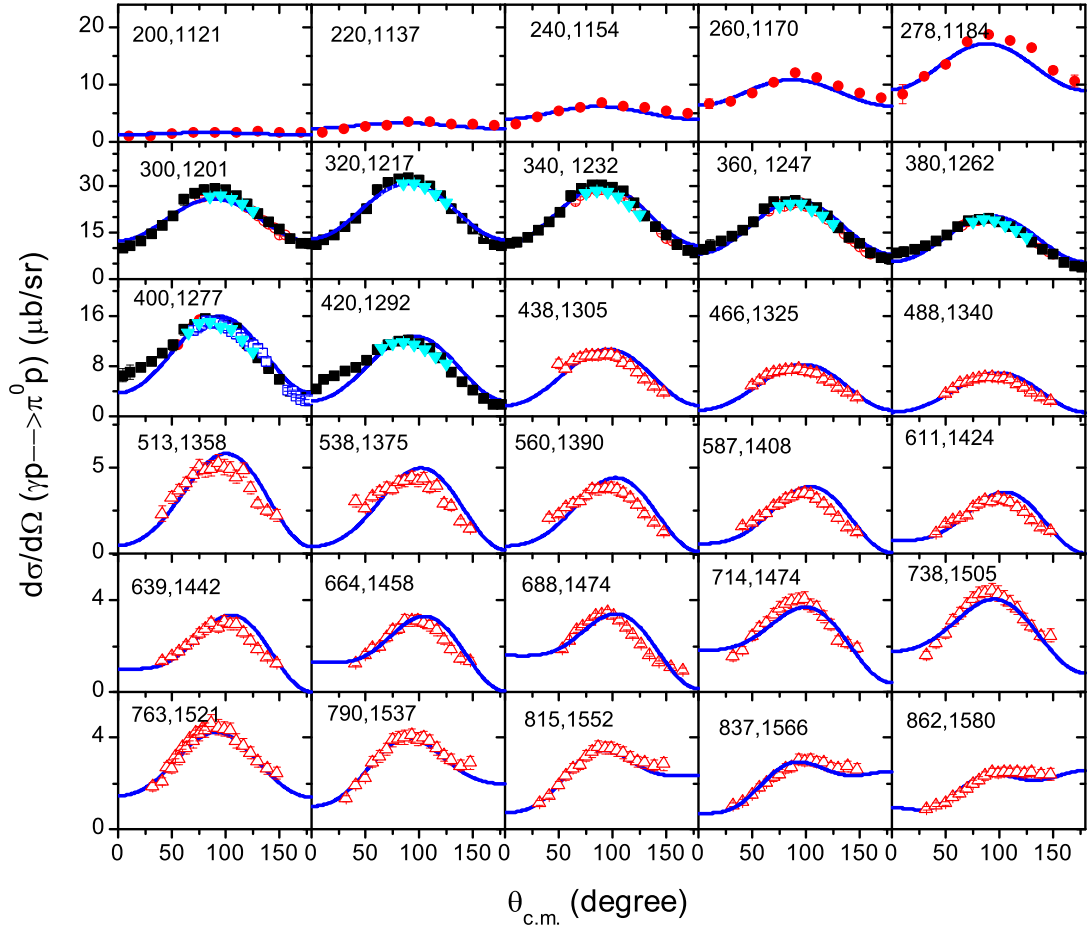} \caption{(Color
online) Differential cross section of the $\gamma p\rightarrow \pi^0
p$ reaction as a function of scattering angle. Data are taken from
\cite{Fuchs:1996ja} (solid circles), \cite{Beck:2006ye} (solid
squares), \cite{Beck:1997ew} (solid triangles), and
\cite{Bartholomy:2004uz} (open triangles). The first and second
numbers in each figure correspond to the photon energy $E_\gamma$
(MeV) and the $\pi N$ center-of-mass (c.m.) energy $W$ (MeV),
respectively.}\label{p1}
\end{figure}
\end{center}
\end{widetext}


The differential and total cross sections are shown in
Figs.~\ref{p1} and ~\ref{p3}, respectively. It is seen that the
chiral quark model can obtain a reasonable description of the data
in a wide energy region $E_\gamma=200\sim 900$ MeV. To clearly see
the contributions from different resonances, we also plot the energy
dependent differential cross sections in Fig.~\ref{p2}. One can
clearly see three bump structures in both the energy dependent
differential cross sections and the total cross section. According
to our calculations, we find that $\Delta(1232)P_{33}$ is
responsible for first bump at $E_\gamma\simeq 300$ MeV. It governs
the reaction in the first resonance region. Both $N(1535)S_{11}$ and
$N(1520)D_{13}$ together dominate the resonance contributions in the
second resonance region. They give approximately equal contributions
to the second bump at $E_\gamma\simeq 700$ MeV. The $N(1720)P_{13}$
resonance might be responsible for the third bump at $E_\gamma\simeq
1000$ MeV. It should be mentioned that although $\Delta(1620)S_{31}$
and $N(1650)S_{11}$ do not give obvious structures in the cross
sections, they are crucial to give a correct shape of the
differential cross sections from the second resonance region to the
third resonance region (see Fig.~\ref{peff}). Switching off their
contributions one can see that the total cross sections around
$E_\gamma= 700\sim 1000$ MeV are overestimated slightly (see
Figs.~\ref{p3}). The $u$-channel background plays a crucial role in
the reaction, it has strong destructive interferences with
$\Delta(1232)P_{33}$, $N(1535)S_{11}$ and $N(1720)P_{13}$. By
including the $t$-channel vector-meson exchange contribution, we
find that the descriptions of the cross sections in the energy
region $E_\gamma= 600\sim 900$ MeV are improved notably, while
without the $t$-channel contributions, the cross sections are
underestimated obviously (see Figs.~\ref{p3} and~\ref{peff}).
Finally, it should be mentioned that our quark model explanation of
the first and second bump structures in the cross sections are
consistent with that of the isobar
model~\cite{vanPee:2007tw,Anisovich:2005tf,Bartholomy:2004uz}.
However, our quark model explanation of the third bump structure
differs from that of the isobar
model~\cite{vanPee:2007tw,Anisovich:2005tf,Bartholomy:2004uz}. In
Ref.~\cite{vanPee:2007tw,Anisovich:2005tf,Bartholomy:2004uz}, the
authors predicted that the third bump might be due to three major
contributions: $\Delta(1700)D_{33}$, $N(1680)F_{15}$ and
$N(1650)S_{11}$, rather than $N(1720)P_{13}$. Thus, to clarify the
puzzle about the third bump structure in the cross section more
studies of the reaction $\gamma p\rightarrow \pi^0 p$ are needed.

\begin{widetext}
\begin{center}
\begin{figure}[ht]
\centering \epsfxsize=17.2 cm \epsfbox{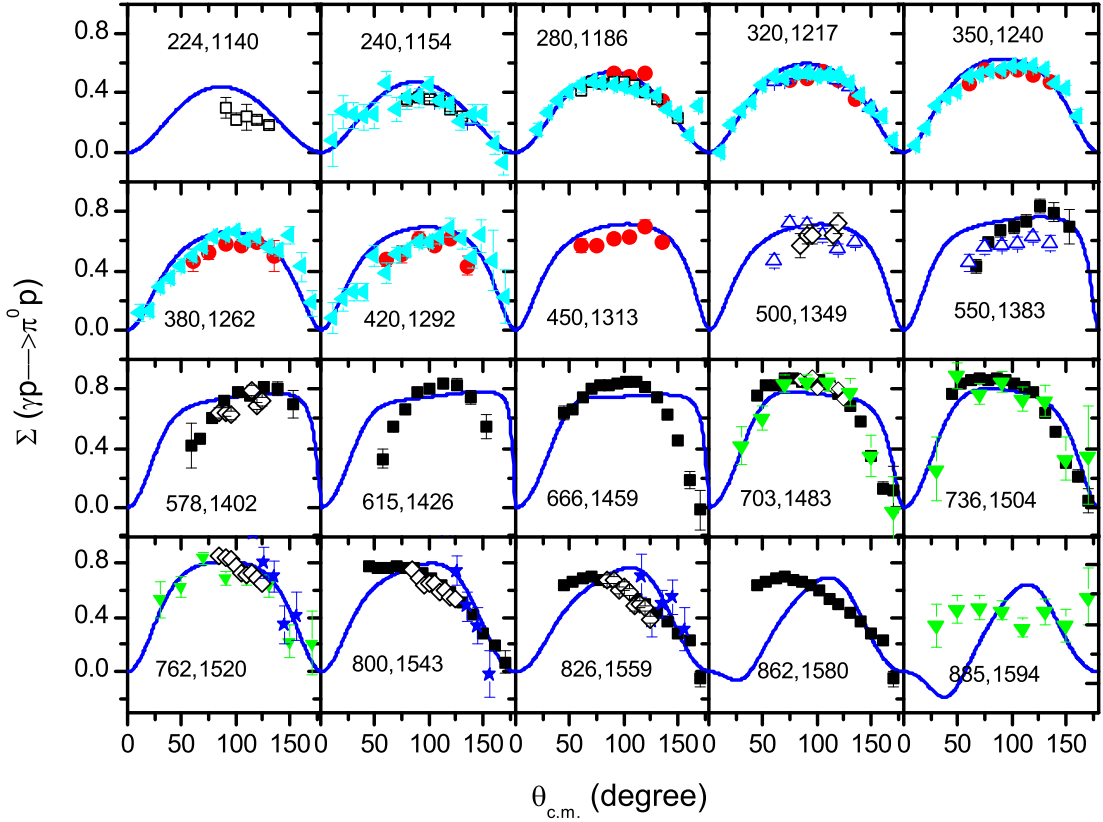} \caption{(Color
online) Beam asymmetry of the $\gamma p\rightarrow \pi^0 p$ process
as a function of scattering angle. The data are taken from
\cite{Beck:2006ye} (solid left triangles), \cite{Blanpied:2001ae}
(open squares), \cite{Ganenko:1974zp} (open up triangles),
\cite{Belyaev:1983xf} (solid circles), \cite{Bartalini:2005wx}
(solid squares), \cite{Adamian:2000yi} (diamonds),
\cite{Knies:1974zx} (solid down triangles), \cite{Elsner:2008sn}
(stars). The first and second number in each figure correspond to
the photon energy $E_\gamma$ (MeV) and the $\pi N$ center-of-mass
energy $W$ (MeV), respectively.}\label{p4}
\end{figure}
\end{center}
\end{widetext}

The beam asymmetries $\Sigma$ in the energy region $E_\gamma=220\sim
900$ MeV are shown in Fig.~\ref{p4}. In this energy region, the
polarized data are not as abundant as those of differential cross
sections. In the low energy region $E_\gamma< 600$ MeV, until now no
data on $\Sigma$ at the forward and backward angles had been
obtained. From Fig.~\ref{p4}, it is seen that the chiral quark model
has achieved good descriptions of the measured beam asymmetries
$\Sigma$ in the energy region $E_\gamma=220\sim 800$ MeV. In the
higher energy region $E_\gamma>800$ MeV, it is found that the chiral
quark model poorly describes the data at forward angles. To clearly
see contributions from different resonances, the energy dependent
beam asymmetries at six angles $\theta_{c. m.}=20^\circ,
60^\circ,90^\circ,125^\circ,150^\circ,170^\circ $ are shown in
Fig.~\ref{p5} as well. From the figure, it is found that the beam
asymmetry $\Sigma$ is sensitive to $\Delta(1232)P_{33}$. Its strong
effects not only exist in the first resonance region, but also
extend to the second resonance region. If we switch off the
contributions of $\Delta(1232)P_{33}$, the beam asymmetry $\Sigma$
changes drastically. Furthermore, we find that both $N(1520)D_{13}$
and $N(1535)S_{11}$ have strong effects on the beam asymmetry
$\Sigma$ around the second resonance region (i.e., $E_\gamma\simeq
700$ MeV), and without their contributions, the beam asymmetry
$\Sigma$ in this energy region changes notably. In the higher energy
region $E_\gamma>800$ MeV, it is found that the resonances
$\Delta(1232)P_{33}$, $N(1520)D_{13}$, $N(1535)S_{11}$,
$N(1650)S_{11}$, $\Delta(1620)S_{31}$ and $N(1720)P_{13}$ together
with the $u$-channel background have equally important contributions
to the beam asymmetry $\Sigma$. It should be mentioned that when the
beam energy $E_\gamma>800$ MeV, many $P$- and $F$-wave states in the
$n=2$ shell begin to have obvious effects on the beam asymmetry
$\Sigma$ as well. Thus, so many equal contributors in this higher
energy region make difficult descriptions of the beam asymmetry
$\Sigma$.

\begin{widetext}
\begin{center}
\begin{figure}[ht]
\centering \epsfxsize=17.2 cm \epsfbox{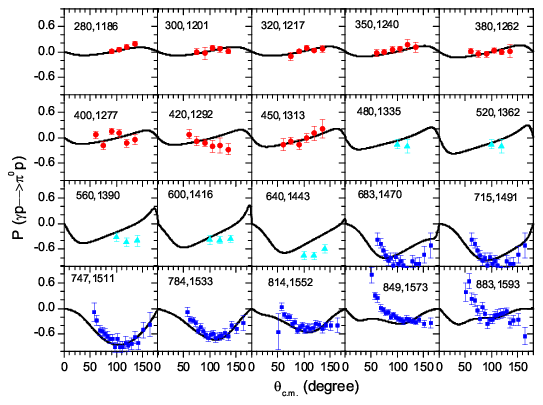} \caption{(Color
online) Polarization of recoil protons $P$ of the $\gamma p
\rightarrow \pi^0p$ reaction as a function of scattering angle. Data
are taken from \cite{Belyaev:1983xf} (circles),
\cite{Bratashevsky:1980pu} (triangles), and \cite{Hartmann:2014mya}
(squares). The first and second number in each figure correspond to
the photon energy $E_\gamma$ (MeV) and the $\pi N$ center-of-mass
energy $W$ (MeV), respectively.}\label{p6}
\end{figure}
\end{center}
\end{widetext}

The polarizations of recoil protons $P$ are shown in Fig.~\ref{p6}.
In the low energy region $E_\gamma<650$ MeV, only a few old data
with limited angle coverage were obtained. Recently, some precise
new data in the higher energy region $E_\gamma\simeq 700\sim 900$
MeV were reported by the CBELSA/TAPS
Collaboration~\cite{Hartmann:2014mya}. From Fig.~\ref{p6}, it is
found that our quark model descriptions are in reasonable agreement
with the measurements in a fairly wide energy region
$E_\gamma=280\sim 800$ MeV. Above the photon energy $E_\gamma\simeq
800$ MeV, the quark model descriptions at both forward and backward
angles become worse compared with the data. To clearly see
contributions from different resonances, the energy dependent $P$ at
six angles $\theta_{c. m.}=40^\circ,
60^\circ,90^\circ,110^\circ,130^\circ,150^\circ $ are shown in
Fig.~\ref{p7} as well. It is found that an obvious dip structure
appears around $E_\gamma=700$ MeV, which can be well described in
the chiral quark model. The dip structure is due to the strong
effects of $\Delta(1232)P_{33}$. When we switch off its
contribution, we find that the dip structure disappears.
Furthermore, from Fig.~\ref{p7} it is obviously seen that the
polarization of recoil protons $P$ is sensitive to $N(1520)D_{13}$
and $N(1535)S_{11}$ around the second resonance region (i.e.,
$E_\gamma\simeq700$ MeV). In the higher energy region $E_\gamma>
800$ MeV, $\Delta(1232)P_{33}$, $N(1520)D_{13}$, $N(1535)S_{11}$,
$N(1650)S_{11}$, the $u$-channel background and other higher partial
waves have approximately equal contributions to $P$, which leads to
a complicated description of the higher energy data.

\begin{widetext}
\begin{center}
\begin{figure}[ht]
\centering \epsfxsize=17.2 cm \epsfbox{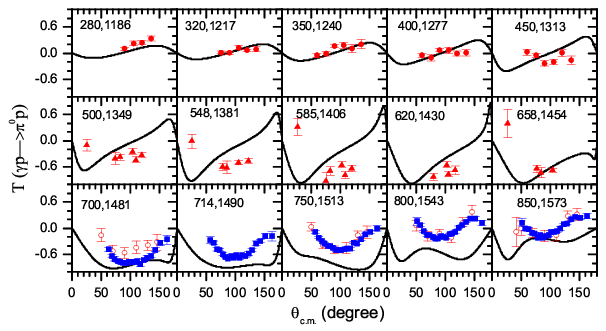} \caption{(Color
online) Target asymmetry of the $\gamma p \to \pi^0p$ reaction as a
function of scattering angle. The data are taken from
\cite{Belyaev:1983xf} (solid circles), \cite{Fukushima:1977xj}
(solid triangles), \cite{Booth:1976es} (open circles),
and~\cite{Hartmann:2014mya} (solid squares). The first and second
number in each figure correspond to the photon energy $E_\gamma$
(MeV) and the $\pi N$ center-of-mass energy $W$ (MeV), respectively.
}\label{p8}
\end{figure}
\end{center}
\end{widetext}

The target asymmetries $T$ are shown in Fig.~\ref{p8}. Below the
photon energy $E_\gamma\simeq 700$ MeV, only a few old data with a
very small angle coverage were obtained. Recently, some precise data
with larger angle coverage in the higher energy region
$E_\gamma\simeq 700\sim 900$ MeV were published by CBELSA/TAPS
Collaboration~\cite{Hartmann:2014mya}. By comparing with the new
data we find that our chiral quark model calculation obviously
underestimates the target asymmetry $T$ in the higher energy region
$E_\gamma> 700$ MeV, but the predicted tendency is in rough
agreement with the data. In the low energy region $E_\gamma\simeq
280\sim 450$ MeV, the data can be well described in the chiral quark
model, though the data at forward and backward angles are still
absent. In the energy region $E_\gamma\simeq 450\sim 660$ MeV, our
quark model results are obviously smaller than the data at the
forward angle. To test our model, we expect that more precise
measurements with large angle coverage can be carried out in the
energy region $E_\gamma<700$ MeV in the future.  To clearly see
contributions from different resonances, the energy dependent target
asymmetries $T$ at six angles $\theta_{c. m.}=27^\circ,
50^\circ,83^\circ,100^\circ,120^\circ,145^\circ $ are shown in
Fig.~\ref{p9} as well. The data show that there is a dip structure
at the angle $\theta_{c. m.}\simeq 80^\circ\sim 100^\circ$ around
the second resonance region $E_\gamma=700$ MeV. This structure can
be explained by the strong interferences between
$\Delta(1232)P_{33}$ and $N(1535)S_{11}$. Switching off the
contributions of either $\Delta(1232)P_{33}$ or $N(1535)S_{11}$, no
obvious dip structure around $E_\gamma=700$ MeV can be found in the
target asymmetry $T$. According to the chiral quark model
predictions, the dip structure should be found at the forward and
backward angles as well. In the higher energy region $E_\gamma> 700$
MeV, it is found that many contributors, such as
$\Delta(1232)P_{33}$, $N(1535)S_{11}$, $N(1650)S_{11}$,
$N(1520)D_{13}$, $\Delta(1620)S_{31}$, $N(1720)P_{13}$ and the
$u$-channel background have obvious effects on the target asymmetry
$T$.


In brief, obvious roles of the $\Delta(1232)P_{33}$,
$N(1535)S_{11}$, $N(1650)S_{11}$, $\Delta(1620)S_{31}$,
$N(1520)D_{13}$ and $N(1720)P_{13}$ have been found in the $\gamma
p\rightarrow \pi^0 p$ process. (i) $\Delta(1232)P_{33}$ not only
plays a dominant role around the first resonance region, its strong
contributions also extend up to the third resonance region, which
can be obviously seen in the cross section, beam asymmetry, target
asymmetry and polarization of recoil protons. (ii) Both
$N(1520)D_{13}$ and $N(1535)S_{11}$ play a dominant role around the
second resonance region. They are the main contributors of the
second bump structure in the energy dependent differential cross
section and total cross section. Their strong effects on the
polarization observables can be seen obviously as well. (iii)
$N(1720)P_{13}$ might play a crucial role in the third resonance
region. It might be responsible for the third bump structure in the
energy dependent differential cross section and total cross section.
However, no dominant role of $N(1720)P_{13}$ is found in the
polarization observables. It should be pointed out that the evidence
of $N(1720)P_{13}$ around the third resonance region should be
further confirmed due to our poor descriptions of the polarization
observables in the higher energy region. (iv) $\Delta(1620)S_{31}$
and $N(1650)S_{11}$ are crucial to give the correct shape of the
differential cross sections in the second resonance region, although
they do not contribute obvious structures in the cross sections. (v)
Furthermore, the $u$- and $t$-channel backgrounds play crucial roles
in the reaction as well. The $u$ channel has a strong interference
with the resonances, such as $\Delta(1232)P_{33}$, $N(1520)D_{13}$
and $N(1535)S_{11}$. Including the $t$-channel vector-meson exchange
contribution, we find that the descriptions in the energy region
$E_\gamma= 600\sim 900$ MeV are improved obviously. (vi) No obvious
contributions of the other resonances, such as $N(1700)D_{13}$,
$\Delta(1700)D_{33}$ and $N(1680)F_{15}$, are found in the $\gamma
p\rightarrow \pi^0 p$ process.

\begin{widetext}
\begin{center}
\begin{figure}[ht]
\centering \epsfxsize=17.2 cm \epsfbox{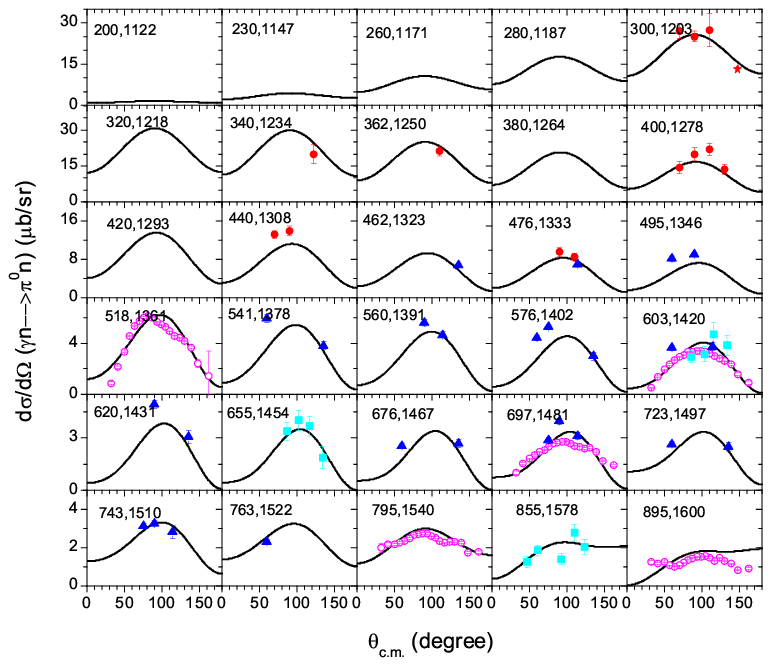} \caption{(Color
online) Differential cross sections of the $\gamma n\rightarrow
\pi^0 n$ reaction as a function of scattering angle. Data are taken
from~\cite{Dieterle:2014blj} (open circles), \cite{Ando:1977} (solid
circles), \cite{Hemmi:1973ii} (solid squares), and
\cite{Bacci:1972bh} (solid triangles). The first and second number
in each figure correspond to the photon energy $E_\gamma$ (MeV) and
the $\pi N$ center-of-mass energy $W$ (MeV),
respectively.}\label{fig-na}
\end{figure}
\end{center}
\end{widetext}

\begin{figure}[ht]
\centering \epsfxsize=8.6 cm \epsfbox{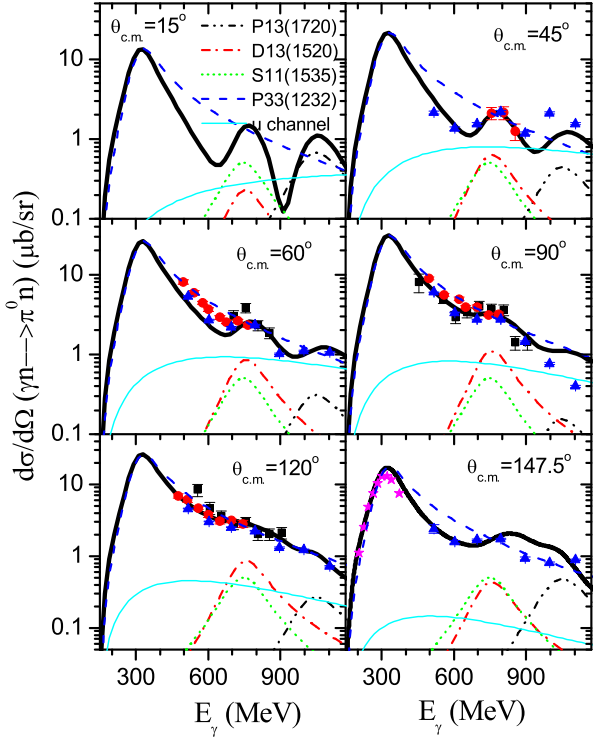} \caption{(Color
online) Photon energy dependent differential cross sections of the
$\gamma n\rightarrow \pi^0 n$ reaction. Data are taken from
~\cite{Dieterle:2014blj}(solid triangles), \cite{Kossert:2003zf}
(solid stars), \cite{Hemmi:1973ii} (solid squares), and
\cite{Bacci:1972bh} (solid circles). The partial cross sections for
$\Delta(1232)P_{33}$, $N(1535)S_{11}$, $N(1520)D_{13}$ and
$N(1720)P_{13}$ are indicated explicitly by different legends in the
figure. }\label{fig-nb}
\end{figure}

\begin{figure}[ht]
\centering \epsfxsize=8.6 cm \epsfbox{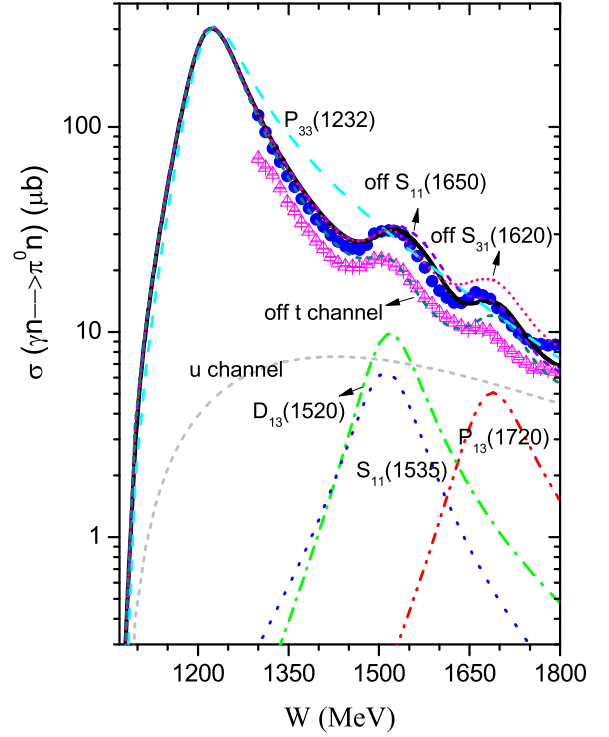} \caption{(Color
online) Total cross section as a function of the c.m. energy $W$ for
the $\gamma n \rightarrow \pi^0 n$ reaction. Data are taken
from~\cite{Dieterle:2014blj}. The results by switching off the
contributions from $N(1650)S_{11}$, $\Delta(1620)S_{31}$, and $t$
channel; and the partial cross sections for $\Delta(1232)P_{33}$,
$N(1535)S_{11}$, $N(1520)D_{13}$, $N(1720)P_{13}$ and $u$ channel
are indicated explicitly by different legends in the
figure.}\label{fig-nc}
\end{figure}

\begin{figure}[ht]
\centering \epsfxsize=8.6 cm \epsfbox{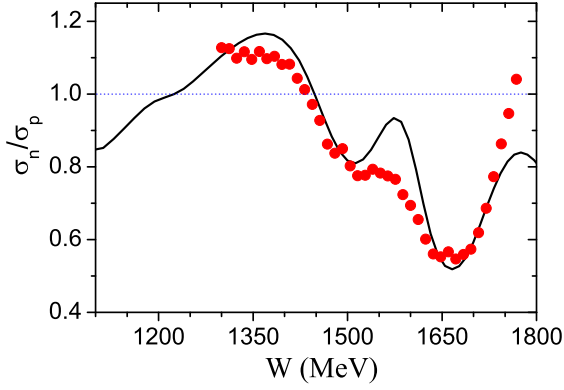} \caption{(Color
online) Cross section ratio $\sigma_n/\sigma_p$ between the
reactions $\gamma n\rightarrow \pi^0 n$ and $\gamma p\rightarrow
\pi^0 p$ as a function of the center-of-mass energy $W$. Data are
taken from Ref.~\cite{Dieterle:2014blj}. }\label{fig-snsp}
\end{figure}

\begin{widetext}
\begin{center}
\begin{figure}[ht]
\centering \epsfxsize=17.2 cm \epsfbox{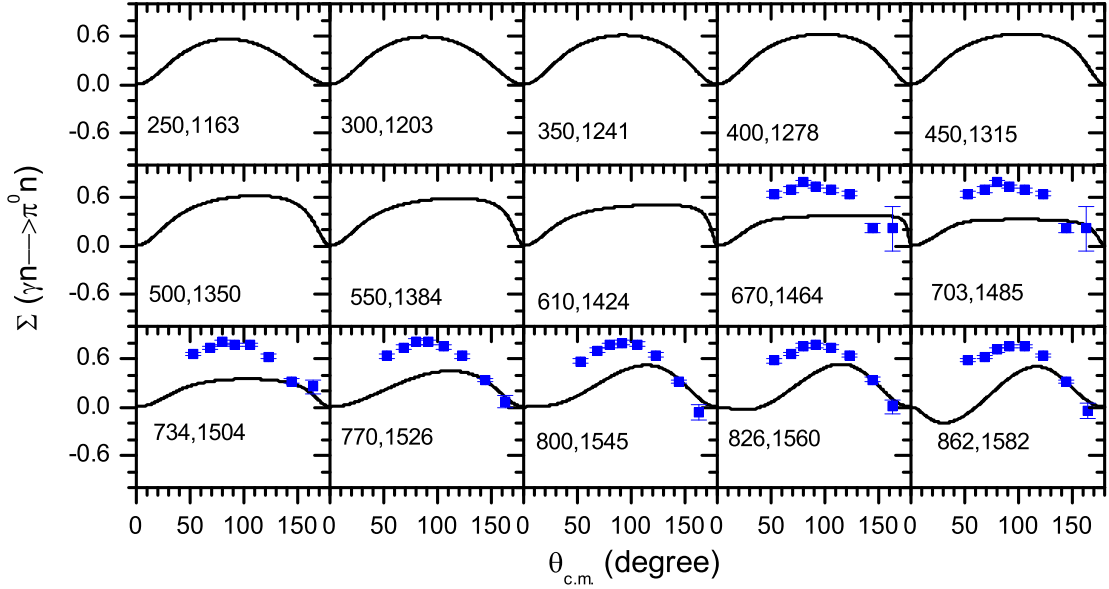} \caption{(Color
online) Beam asymmetry of the $\gamma n\rightarrow \pi^0 n$ reaction
as a function of scattering angle. Data are taken
from~\cite{DiSalvo:2009zz}. The first and second numbers in each
figure correspond to the photon energy $E_\gamma$ (MeV) and the $\pi
N$ center-of-mass energy $W$ (MeV), respectively.}\label{fig-nd}
\end{figure}
\end{center}
\end{widetext}

\begin{figure}[ht]
\centering \epsfxsize=8.6 cm \epsfbox{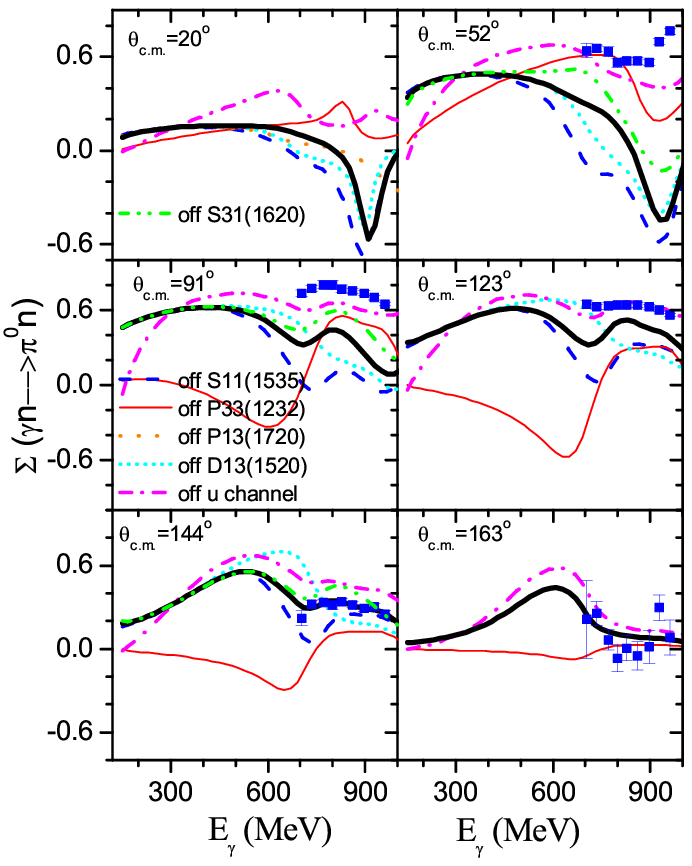} \caption{(Color
online) Photon energy dependent beam asymmetry of the $\gamma n
\rightarrow \pi^0n$ reaction.  Data are taken
from~\cite{DiSalvo:2009zz}. The results by switching off the
contributions from various partial waves are indicated explicitly by
different legends in the figure. }\label{fig-ne}
\end{figure}

\subsection{$\gamma n\rightarrow \pi^0 n$}

The chiral quark model studies of $\gamma n\rightarrow \pi^0 n$ were
carried out in Refs.~\cite{Li:1994cy,Li:1997gd,Zhao:2002id} about
twenty years ago. However, the model studies were limited in the
first resonance region, because only a few scattered data were
obtained from the old measurements in the early 1970s. Fortunately,
obvious progress has been achieved in experiments in recent years.
In 2009, some measurements of the beam asymmetries for the $\gamma
n\rightarrow \pi^0 n$ process were obtained by the GRAAL experiment
in the second and third resonances region~\cite{DiSalvo:2009zz}. In
this energy region, recently the quasi-free differential and total
cross sections for this reaction were also measured by the Crystal
Ball/TAPS experiment at MAMI~\cite{Dieterle:2014blj}. Thus, these
new measurements in the higher resonances region provide us a good
opportunity to extend the chiral quark model to study these
high-lying resonances.

The contributors of the $s$-channel intermediate states classified
in the quark model with $n\leq 2$ have been listed in
Tab.~\ref{CGLN2}. In the $n=0$ shell, the dominant contribution to
the reaction comes from the $\Delta(1232)P_{33}$, which has much
larger CGLN amplitudes than the nucleon pole. In the $n=1$ shell,
three $S$-wave states $N(1535)S_{11}$, $N(1650)S_{11}$ and
$\Delta(1620)S_{31}[70,^210]$, and four $D$-wave states
$N(1520)D_{13}$, $N(1700)D_{13}$, $N(1675)D_{15}$ and
$\Delta(1700)D_{33}$ contribute to the reaction. By comparing their
CGLN amplitudes listed in Tab.~\ref{CGLN2}, we find that
$N(1535)S_{11}$ and $N(1520)D_{13}$ play dominant roles in these
$S$- and $D$-wave resonances. In the $n=2$ shell, twelve $P$-wave
resonance and seven $F$-wave resonances contribute to the reaction.
Most of the $P$-wave and $F$-wave resonances in the $n=2$ shell have
comparable amplitudes. $N(1720)P_{13}$, $N(1900)P_{13}$ and
$\Delta(1600)P_{33}$ have relatively larger CGLN amplitudes in the
$n=2$ shell $P$-wave resonances, while $N(1680)F_{15}$ and
$\Delta(1905)F_{35}$ have relatively bigger CGLN amplitudes among
the $n=2$ shell $F$-wave resonances.

In this work, we have carried out a chiral quark model study of the
$\gamma n\rightarrow \pi^0 n$ reaction up to the second and third
resonances region. In the SU(6)$\otimes$O(3) symmetry limit, the
parameters from the $u$- and $t$-channel backgrounds and the
$\Delta$ resonances $\Delta(1232)P_{33}$ and $\Delta(1620)S_{31}$
for the $\pi^0 n$ channel should be the same as those for the $\pi^0
p$ channel, which have been well determined by the $\gamma p$ data.
Thus, in the $\gamma n\rightarrow \pi^0 n$ reaction these parameters
are taken to have the same values as in the $\gamma p\rightarrow
\pi^0 p$ process. The other strength parameters, $C_R$, for the main
resonances $N(1535)S_{11}$, $N(1650)S_{11}$, $N(1520)D_{13}$ and
$N(1720)P_{13}$ for the $\gamma n$ reaction can not be well
constrained by the $\gamma p$ data for their different
photocouplings, thus, we determine them by fitting the 36 $\gamma n$
data points of total cross section around the second resonance
energy region 1.30 GeV$\leq W \leq 1.72$ GeV recently measured at
MAMI~\cite{Dieterle:2014blj}. The $\chi^2$ datum point is about
$\chi^2/N_{data}=2.8$. Our results compared with the data have been
shown in Figs~\ref{fig-na}-\ref{fig-ne}.

The differential cross sections compared with the data are shown in
Fig.~\ref{fig-na}. In the energy region what we consider, only a few
data can be obtained. Fortunately, the abundant data for the $\gamma
p\to \pi^0 p$ process help us well constrain some important model
parameters, as we pointed out above. From Fig.~\ref{fig-na}, one can
see that the data of the $\gamma n\rightarrow \pi^0 n$ reaction are
reasonably reproduced. To clearly see the contributions from
different partial waves, we plot the energy dependent differential
cross sections in Fig.~\ref{fig-nb} as well. Our results obviously
show three bump structures in the forward angle region. It is found
that the resonances $\Delta(1232)P_{33}$, $N(1535)S_{11}$,
$N(1520)D_{13}$ and $N(1720)P_{13}$ play crucial roles in the
$\gamma n\rightarrow \pi^0 n$ reaction. The $\Delta(1232)P_{33}$
resonance is responsible for the first bump structure around
$E_\gamma\simeq 300$ MeV. Both $N(1535)S_{11}$ and $N(1520)D_{13}$
are the main contributors to the second bump around $E_\gamma\simeq
700$ MeV. The $N(1720)P_{13}$ resonance is most likely responsible
for the third bump around $E_\gamma\simeq 1000$ MeV.

The total cross sections compared with the data are shown in
Fig.~\ref{fig-nc}. Obvious roles of $\Delta(1232)P_{33}$,
$N(1535)S_{11}$ and $N(1720)P_{13}$ in the $\gamma n\rightarrow
\pi^0 n$ reaction can be found in the total cross section as well.
Recently, the total cross section was measured by the Crystal
Ball/TAPS experiment at MAMI~\cite{Dieterle:2014blj}. There are two
obvious bump structures in the cross section in the second and third
resonances region (see Fig.~\ref{fig-nc}). The bump structure around
the second resonance region receives approximately equal
contributions from $N(1535)S_{11}$ and $N(1520)D_{13}$, while the
bump structure around the third resonance region might be due to the
contributions of $N(1720)P_{13}$. There are no measurements of the
total cross section in the first resonance region. In this energy
region, we predict that the ratio of total cross section between the
$\pi^0 n$ channel and the $\pi^0 p$ channel $\sigma_n/\sigma_p$ is
around $1$ (see Fig.~\ref{fig-snsp}).

Furthermore, by analyzing the data of differential and total cross
sections, we find that $\Delta(1620)S_{31}$ and $N(1650)S_{11}$ play
obvious roles around their mass threshold. If we switch off them,
the cross sections around their mass threshold are overestimated
significantly. It should be mentioned that, the role of
$N(1650)S_{11}$ should be confirmed by more accurate data in the
future, which will be further discussed in Sec.~\ref{Hel}. Finally,
it should be pointed out that the backgrounds play a crucial role in
the reaction. The $u$-channel background has strong destructive
interferences with $\Delta(1232)P_{33}$, $N(1535)S_{11}$,
$N(1520)D_{13}$ and $N(1720)P_{13}$. Including the $t$-channel
vector-meson exchange contribution, we find that the descriptions of
the cross sections in the energy region $E_\gamma= 600\sim 900$ MeV
are improved significantly.

The polarization observations for the $\gamma n\rightarrow \pi^0 n$
reaction are very sparse. In the year of 2009, the beam asymmetry
$\Sigma$ in the second and third resonances region was measured by
the GRAAL Collaboration for the first time ~\cite{DiSalvo:2009zz}.
Our chiral quark model results are shown in Fig.~\ref{fig-nd}. From
the figure, it is seen that the model results are in rough agreement
with the data. Our results are notably smaller than the data at the
intermediate angles. To clearly see the contributions from different
partial waves, the energy dependent beam asymmetries $\Sigma$ at six
angles $\theta_{c. m.}=20^\circ,
52^\circ,91^\circ,123^\circ,144^\circ,163^\circ $ are shown in
Fig.~\ref{fig-ne} as well. From the figure, one can find that below
the photon energy $E_\gamma\simeq 500$ MeV, the beam asymmetry is
sensitive to $\Delta(1232)P_{33}$ and the $u$-channel background. By
turning off one of them, the beam asymmetry changes drastically in
this energy region. Similarly, we can obviously find that around the
second resonance region, i.e., $E_\gamma\sim700$ MeV, the
$N(1535)S_{11}$, $N(1650)S_{11}$, $N(1520)D_{13}$,
$\Delta(1232)P_{33}$, $\Delta(1620)S_{31}$ and the $u$-channel
background have strong effects on the beam asymmetry. Up to the
second resonance region, the higher partial wave states, such as
$N(1720)P_{13}$ begin to contribute to beam asymmetry. Many
resonances together with the backgrounds have approximately equal
contributions to the beam asymmetry, leading to a very complicated
description of the data.

As a whole, a reasonable chiral quark model description of the
$\gamma n\rightarrow \pi^0 n$ reaction is obtained from the pion
production threshold up to the second resonance region. Obvious
evidences of the $\Delta(1232)P_{33}$, $N(1535)S_{11}$,
$N(1520)D_{13}$ and $N(1720)P_{13}$ are also found in the $\gamma
n\rightarrow \pi^0 n$ reaction. (i) The ground state
$\Delta(1232)P_{33}$, the $S$-wave state $N(1535)S_{11}$ together
with the $D$-wave state $N(1520)D_{13}$, and the $P$-wave state
$N(1720)P_{13}$ are responsible for the first, second, and third
bump structures in the cross sections, respectively. (ii)
Furthermore, another two $S$-wave states $\Delta(1620)S_{31}$ and
$N(1650)S_{11}$ have obvious effects on the differential cross
section around their mass threshold, although they do not give any
structure in the cross sections. It should be pointed out that the
role of $N(1650)S_{11}$ should be further confirmed in future
experiments. (iii) The backgrounds play a crucial role in the
reaction. The $u$ channel background has a strong constructive
interference with the $s$-channel resonances $\Delta(1232)P_{33}$,
$N(1535)S_{11}$ and $N(1520)D_{13}$. By including the $t$-channel
vector-meson exchange contribution, we find that the descriptions in
the energy region $E_\gamma= 600\sim 900$ MeV are slightly improved.
(vi) No obvious evidence of the other resonances, such as
$N(1700)D_{13}$, $N(1675)D_{15}$, $\Delta(1700)D_{33}$ and
$N(1680)F_{15}$, is found in the $\gamma n\rightarrow \pi^0 n$
process.

\begin{table}[ht]
\caption{The expressions of $\xi$ in Eq.~\ref{hlc} for various
resonances. Here we have defined
$\mathcal{K}\equiv\sqrt{\frac{\alpha_e\alpha_\pi (E_f+M_N)\pi}{2
M_R^2M_N}}\frac{1}{\Gamma_R}$, $\mathcal{A}\equiv
\left[\frac{2\omega_\gamma}{m_q}-\frac{2q^2}{3\alpha^2}(1+\frac{\omega_\pi}{E_f+M_N})
\right]e^{-\frac{k^2+q^2}{6\alpha^2}}$,
$\mathcal{B}\equiv\frac{2q^2}{3\alpha^2}(1+\frac{\omega_\pi}{E_f+M_N})e^{-\frac{k^2+q^2}{6\alpha^2}}$,
$\mathcal{D}\equiv\left[\frac{2\omega_\gamma}{m_q}-\frac{2q^2}{5\alpha^2}(1+\frac{\omega_\pi}{E_f+M_N})\right]e^{-\frac{k^2+q^2}{6\alpha^2}}$.
}\label{hlx}
\begin{tabular}{|c c|c| }\hline
$\Delta(1232)P_{33}$ & $\xi_{1/2}$ &
$-\mathcal{K}\sqrt{\frac{1}{2}}\frac{4\omega_\gamma}{9m_q}(1+\frac{\omega_\pi}{E_f+M_N})|\mathbf{q}|C_{P_{33}(1232)}$     \\
                       & $\xi_{3/2}$ & $-\mathcal{K}\sqrt{\frac{3}{2}}\frac{4\omega_\gamma}{9m_q}(1+\frac{\omega_\pi}{E_f+M_N})|\mathbf{q}|C_{P_{33}(1232)}$
                       \\
\hline
 $\Delta(1620)S_{31}$ & $\xi_{1/2}$ &
$\mathcal{K}\frac{\omega_\gamma}{18}(1-\frac{\omega_\gamma}{6m_q})\mathcal{A}
C_{S_{31}(1620)}$     \\
 \hline $N(1535)S_{11}$ & $\xi^p_{1/2}$ &
$\mathcal{K}\frac{\omega_\gamma}{9}(1+\frac{\omega_\gamma}{2m_q})\mathcal{A}
C^{[70,^28]}_{S_{11}(1535)}$     \\
                       & $\xi^n_{1/2}$ & $-\mathcal{K}
\frac{\omega_\gamma}{9}\left[(1+\frac{\omega_\gamma}{6m_q})+\frac{\tan\theta_S\omega_\gamma}{6m_q}
\right]\mathcal{A}C^{[70,^28]}_{S_{11}(1535)}$ \\
\hline $N(1650)S_{11}$  & $\xi^p_{1/2}$ &
$\mathcal{K}\frac{\omega_\gamma}{9}(1+\frac{\omega_\gamma}{2m_q})\mathcal{A}C^{[70,^28]}_{S_{11}(1650)}$      \\
                &  $\xi^n_{1/2}$ & $-\mathcal{K}
\frac{\omega_\gamma}{9}\left[(1+\frac{\omega_\gamma}{6m_q})
-\frac{\cot\theta_S\omega_\gamma}{6m_q}
\right]\mathcal{A}C^{[70,^28]}_{S_{11}(1650)}$     \\
\hline
        $N(1520)D_{13}$     & $\xi^p_{1/2}$ & $\mathcal{K}
\frac{\omega_\gamma}{9\sqrt{2}}(1-\frac{\omega_\gamma}{m_q})
\mathcal{B}C^{[70,^28]}_{D_{13}(1520)}
$    \\
               &  $\xi^p_{3/2}$ & $\mathcal{K} \sqrt{\frac{3}{2}}\frac{\omega_\gamma}{9}
\mathcal{B}C^{[70,^28]}_{D_{13}(1520)}$    \\
              &  $\xi^n_{1/2}$ & $-\mathcal{K}
\frac{\omega_\gamma}{9\sqrt{2}}\left[(1-\frac{\omega_\gamma}{3m_q})
-\frac{\tan\theta_D\omega_\gamma}{3\sqrt{10}m_q}
\right]\mathcal{B}C^{[70,^28]}_{D_{13}(1520)}$    \\
             &  $\xi^n_{3/2}$ & $-\sqrt{\frac{3}{2}}\mathcal{K}
\frac{\omega_\gamma}{9}\left[
1-\frac{\tan\theta_D\omega_\gamma}{\sqrt{10}m_q}
\right]\mathcal{B}C^{[70,^28]}_{D_{13}(1520)}$
\\ \hline
        $N(1700)D_{13}$  & $\xi^n_{1/2}$&
$-\frac{\omega_\gamma}{9\sqrt{2}}\mathcal{K}
\left[(1-\frac{\omega_\gamma}{3m_q})
+\frac{\cot\theta_D\omega_\gamma}{3\sqrt{10}m_q}
\right]\mathcal{B}C^{[70,^28]}_{D_{13}(1700)}$\\
             &  $\xi^n_{3/2}$ & $-\sqrt{\frac{3}{2}}\mathcal{K}
\frac{\omega_\gamma}{9}\left[
1+\frac{\cot\theta_D\omega_\gamma}{\sqrt{10}m_q}
\right]\mathcal{B}C^{[70,^28]}_{D_{13}(1700)}$    \\
             &  $\xi^p_{1/2}$ & $\mathcal{K}
\sqrt{\frac{1}{2}}\frac{\omega_\gamma}{9}(1-\frac{\omega_\gamma}{m_q})
\mathcal{B}C^{[70,^28]}_{D_{13}(1700)}
$ \\
             &  $\xi^p_{3/2}$ & $\mathcal{K} \sqrt{\frac{3}{2}}\frac{\omega_\gamma}{9}
\mathcal{B}C^{[70,^28]}_{D_{13}(1700)}$
\\ \hline
$N(1675)D_{15}$&  $\xi^n_{1/2}$ & $-\mathcal{K}
\frac{\omega_\gamma^2}{40m_q} \mathcal{B}C_{D_{15}(1675)}
$    \\
            &  $\xi^n_{3/2}$ & $-\mathcal{K} \frac{\omega_\gamma^2}{20\sqrt{2}m_q}
\mathcal{B}C_{D_{15}(1675)}
$    \\
\hline $N(1720)P_{13}$&  $\xi^p_{1/2}$ & $\mathcal{K}
\sqrt{\frac{1}{2}}\frac{5}{108}\frac{\omega_\gamma^2}{\alpha^2}
(1+\frac{k}{3m_q})\mathcal{D}|\mathbf{q}|C^p_{P_{13}(1720)}
$    \\
            &  $\xi^p_{3/2}$ & $-\mathcal{K}
\sqrt{\frac{1}{6}}\frac{5}{108}\frac{\omega_\gamma^2}{\alpha^2}
\mathcal{D}|\mathbf{q}|C^p_{P_{13}(1720)}
$    \\
 & $\xi^n_{1/2}$ & $-\mathcal{K}
\sqrt{\frac{1}{2}}\frac{5}{108}\frac{\omega_\gamma^2}{\alpha^2}
\frac{2k}{27m_q}\mathcal{D}|\mathbf{q}|C^n_{P_{13}(1720)}
$    \\
            &  $\xi^n_{3/2}$ & 0    \\
\hline
\end{tabular}
\end{table}

\subsection{Helicity amplitudes}\label{Hel}

The accurate data for the $\gamma n\rightarrow \pi^0 n$ and $\gamma
p\rightarrow \pi^0 p$ processes provide us a good platform to
extract the helicity amplitudes of the dominant resonances in these
reactions. Theoretically, the helicity amplitudes $A_\lambda$ for a
baryon resonance $N^*$ photoexcitation on a nucleon are defined by
\begin{eqnarray}
A_{\lambda}=\sqrt{2\pi/k}\langle N^*; J_z=\lambda |H_e|N;
J_z=\lambda -1\rangle,
\end{eqnarray}
where $\lambda=1/2$ and $3/2$. As we know, the helicity amplitudes
of a resonance are related to the transition amplitudes of the
photoproduction reactions. Thus, we can extract the helicity
amplitudes from the neutral pion photoproduction processes by the
relation
\begin{eqnarray}\label{hlc}
A^{n,p}_{1/2,3/2}=\sqrt{\frac{|\mathbf{q}|M_R\Gamma_R}{|\mathbf{k}|M_Nb_{\pi^0N}}}\xi^{n,p}_{1/2,3/2},
\end{eqnarray}
where $b_{\pi^0N}\equiv \Gamma_{\pi^0N}/\Gamma_R$ is the branching
ratio of the resonance. The quantity $\xi$ for different resonances
can be analytically expressed from their CGLN amplitudes. We have
given the expressions of the $\xi$ for several low-lying nucleon and
$\Delta$ resonances in Tab.~\ref{hlx}. We estimate the helicity
amplitudes for these main contributing resonances:
$\Delta(1232)P_{33}$, $\Delta(1620)S_{31}$, $N(1535)S_{11}$,
$N(1650)S_{11}$, $N(1520)D_{13}$ and $N(1720)P_{13}$. The branching
ratios $b_{\pi N}$ for $N(1720)P_{13}$ are adopted from our quark
model prediction, and the branching ratios for other resonances are
taken from PDG14~\cite{Agashe:2014kda} (see table \ref{Br}). Our
extracted helicity amplitudes are listed in Tab.~\ref{HL}. As a
comparison, in the same table we also show our previous solution
extracted from the $\eta$ photoproduction
processes~\cite{Zhong:2011ti}, the recent analysis of the $\gamma N$
data from SAID~\cite{Workman:2012jf,Chen:2012yv,Workman:2011vb},
Kent~\cite{Shrestha:2012ep} and
BnGa~\cite{Anisovich:2009zy,Anisovich:2013jya}, the average values
from PDG14~\cite{Agashe:2014kda}, and the theoretical predictions
from different quark models~\cite{Li:1990qu,Capstick:1992uc}.

\begin{widetext}
\begin{center}
\begin{table}[ht]
\caption{Branching ratio $b_{\pi N}$ of the resonances used in the
calculation.} \label{Br}
\begin{tabular}{|c|c|c|c|c|c|c|c|c|c|c|c|c|c| }\hline\hline
 Resonance&$\Delta(1232)P_{33}$&$\Delta(1620)S_{31}$&$N(1535)S_{11}$&$N(1650)S_{11}$&$N(1520)D_{13}$&$N(1720)P_{13}$\\ \hline
 $b_{\pi N}$& 1.0 &$20\sim30\%$&$35\sim 55\%$&$50\sim 90\%$&$55\sim 65\%$&$60\sim 90\%$\\ \hline
\end{tabular}
\end{table}
\end{center}
\end{widetext}

\begin{widetext}
\begin{center}
\begin{table}[ht]
\caption{Extracted helicity amplitudes for the main nucleon and
$\Delta(1232)$ resonances from the neutral pion photoproduction
reactions (in units of $ 10^{-3}$GeV$^{-1/2}$).} \label{HL}
\begin{tabular}{|c|c|c|c|c|c|c|c|c|c|c|c|c|c| }\hline\hline
resonance& helicity & this work & ZZ11\cite{Zhong:2011ti}
 &PDG14~\cite{Agashe:2014kda}
&Kent12~\cite{Shrestha:2012ep} &
BnGa~\cite{Anisovich:2009zy,Anisovich:2013jya}
 & SAID12~\cite{Workman:2012jf,Chen:2012yv} & SAID11\cite{Workman:2011vb}&ZF \cite{Li:1990qu} & C92 \cite{Capstick:1992uc}\\\hline
$\Delta(1232)P_{33}$  & $A_{1/2}$    &$-133$      &   ...        & $-135\pm6$  &$-137\pm1$  & $-136\pm5$ & $-139\pm2$&$-138\pm3$&$-94$ &$-108$\\
                      & $A_{3/2}$    &$-230$      &   ...        & $-255\pm 5$ &$-251\pm1$  & $-267\pm8$ &$-262\pm3$ &$-259\pm5$&$-162$& $-186$\\\hline
$N(1535)S_{11}$       & $A_{1/2}^p$  &$137\pm 15$ &$60\pm 5$  & $115\pm15$  &$59\pm 3$   & $90\pm15$  &$128\pm4$  &$99\pm2$  &142   & 76 \\
                      & $A_{1/2}^n$  &$-77\pm 9$  &$-68\pm 5$ & $-75\pm20$  &$-49\pm3$   & $-93\pm11$ &$-58\pm6$  &$-60\pm3$ &$-77$ & $-63$ \\\hline
$N(1650)S_{11}$       & $A_{1/2}^p$  &$61\pm 9$   &$41\pm 13$ & $45\pm10$   &$30\pm 3$   &$60\pm20$   &$55\pm30$  &$65\pm25$ &$78$  & $54$ \\
                      & $A_{1/2}^n$  &$-18\pm 3$   &$24\pm 7$  & $-50\pm20$  &$11\pm 2$   &$25\pm20$   &$-40\pm10$ &$-26\pm8$ &$-47$ & $-35$\\\hline
$\Delta(1620)S_{31}$  & $A_{1/2}$    &$80\pm 8$  &           & $40\pm15$   &$-3\pm 3$   &$63\pm12$   &$29\pm3$   &$64\pm2$  &$72$  & 81\\
\hline
$N(1520)D_{13}$       & $A_{1/2}^p$  &$-17\pm 1$  &$-32\pm 7$ & $-20\pm 5$  &$-34\pm1$   &$-32\pm 6$  & $-19\pm 2$&$-16\pm 2$&$-47$ &$-15$ \\
                      & $A_{3/2}^p$  &$109\pm 5$   &$113\pm 23$& $140\pm 10$ &$127\pm3$   &$138\pm 8$  & $141\pm 2$&$156\pm 2$&117   &$134$\\
                      & $A_{1/2}^n$  &$-30\pm 1$  &$-40\pm8$  & $-50\pm10$  &$-38\pm3$   &$-49\pm8$   & $-46\pm6$ &$-47\pm2$ &$-75$ &$-38$\\
                      & $A_{3/2}^n$  &$-90\pm 4$  &$-126\pm 26$& $-115\pm10$ &$-101\pm4$ &$-113\pm12$ & $-115\pm5$&$-125\pm2$&$-127$&$-114$ \\\hline
$N(1720)P_{13}$       & $A_{1/2}^p$  &$-89\pm 9$  &  ...         & $100\pm20$   &$57\pm3$   &$130\pm50$  &$95\pm 2$  &$99\pm 3$ &$-68$ & $-11$\\
                      & $A_{3/2}^p$  &$34\pm 4$   &  ...         & ...             &$-19\pm2$  &$100\pm50$  &$-48\pm 2$ &$-43\pm 2$&$53$  & $-31$ \\
                      & $A_{1/2}^n$  &$18\pm2$    &  ...         & ...             &$-2\pm1$   &$-80\pm50$  &  ...         &$-21\pm 4$&$-4$  & $4$\\
                      & $A_{3/2}^n$  &0           &  ...         & ...            &$-1\pm2$   &$-140\pm65$ &   ...        &$-38\pm 7$&$-33$ & $11$  \\\hline
\hline
\end{tabular}
\end{table}
\end{center}
\end{widetext}

From Tab.~\ref{HL}, it is found that the helicity amplitudes of
$\Delta(1232)P_{33}$ extracted in present work are in good agreement
with the values from PDG14~\cite{Agashe:2014kda} and other partial
wave analysis
groups~\cite{Shrestha:2012ep,Anisovich:2009zy,Anisovich:2013jya,Workman:2012jf,Chen:2012yv,Workman:2011vb,Cao:2013psa}.

The $A^{p}_{1/2}$ and $A^{n}_{1/2}$ of $N(1535)S_{11}$ extracted in
this work are compatible with the PDG average values and the latest
analysis of the $\gamma N$ data from the
SAID~\cite{Workman:2012jf,Chen:2012yv,Workman:2011vb}. It should be
pointed out that with the same model we found a smaller $\gamma p$
coupling $A^{p}_{1/2}\simeq 60\times10^{-3}$GeV$^{-1/2}$ for the
$N(1535)S_{11}$ by the analysis of the $\gamma p\to \eta p$
process~\cite{Zhong:2011ti}, and similar solution was also obtained
in~\cite{He:2008ty,Shklyar:2012js}. The reason for the different
$\gamma p$ couplings for $N(1535)S_{11}$ in the $\pi^0 p$ and $\eta
p$ channels should be clarified in future studies.

\begin{figure}[ht]
\centering \epsfxsize=8.6 cm \epsfbox{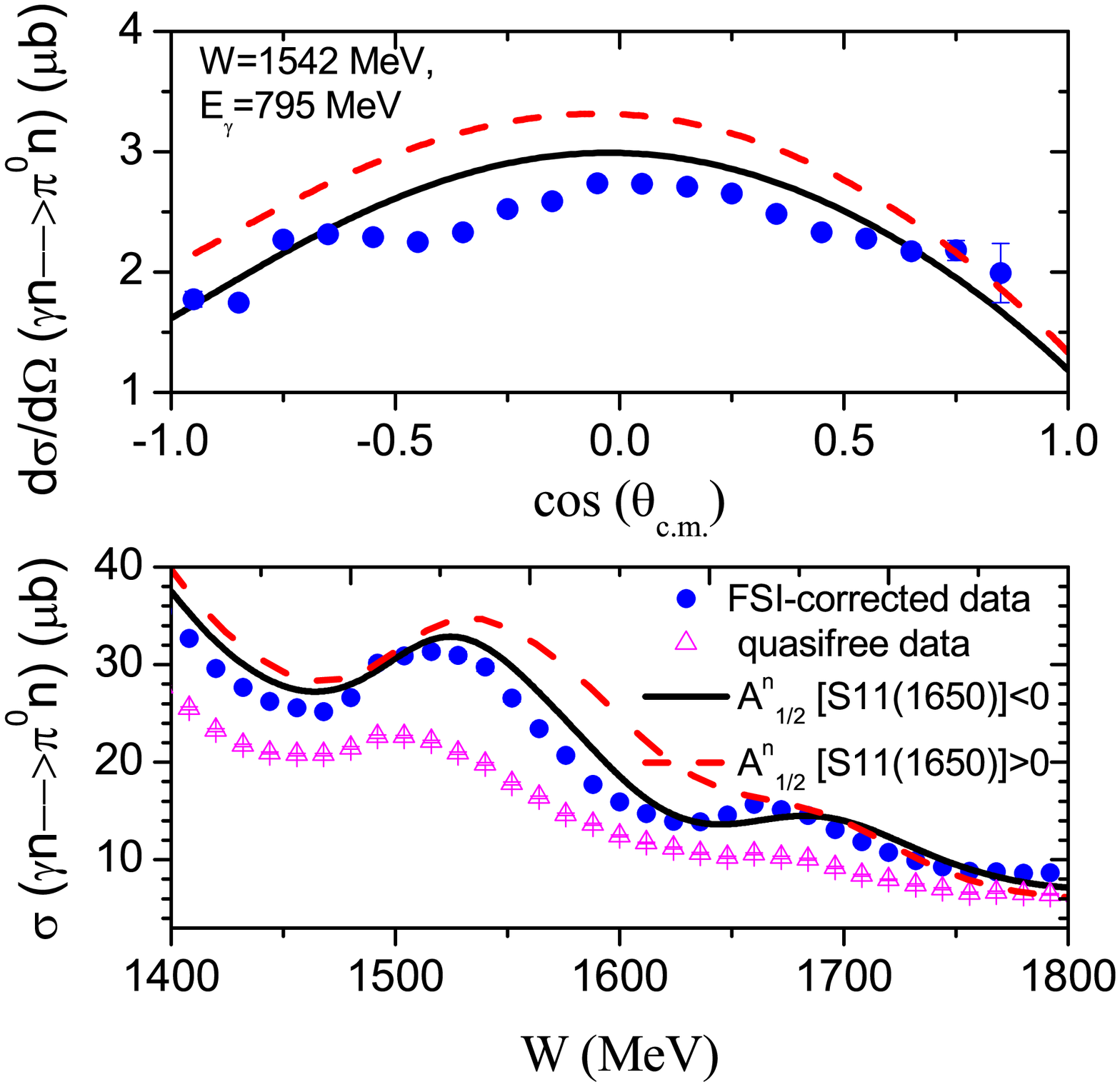} \caption{(Color
online) Effects of $N(1650)S_{11}$ on the differential cross
sections and the total cross section around its mass threshold. Data
are taken from Ref.~\cite{Dieterle:2014blj}. The solid and dashed
curves are for the results with negative and positive $\gamma n$
couplings for $N(1650)S_{11}$, respectively. }\label{fig-eff}
\end{figure}

All the partial wave analysis groups have extracted similar $\gamma
p$ coupling $A^{p}_{1/2}$ for $N(1650)S_{11}$ from the data, which
is also consistent with the theoretical predictions in quark
models~\cite{Li:1990qu,Capstick:1992uc}. However, contradictory
solutions for the $\gamma n$ coupling $A^{n}_{1/2}$ of
$N(1650)S_{11}$ are obtained by different groups. Our previous
analysis of the $\gamma n\to \eta n$ reaction indicates a positive
helicity coupling $A^{n}_{1/2}\simeq 24\times10^{-3}$GeV$^{-1/2}$
for $N(1650)S_{11}$~\cite{Zhong:2011ti}, which is supported by the
latest analysis of the same reaction from the
BnGa~\cite{Anisovich:2013jya,Anisovich:2015tla} and
Kent~\cite{Shrestha:2012ep} groups. However, in present work by
analyzing the recent the final-state-interaction (FSI) corrected
data of the $\gamma n\to \pi^0 n$ reaction from the A2
Collaboration~\cite{Dieterle:2014blj}, a negative helicity coupling
$A^{n}_{1/2}\simeq -18 \times10^{-3}$GeV$^{-1/2}$ is obtained, which
is compatible with the values from the PDG14~\cite{Agashe:2014kda}
and the recent SAID
analysis~\cite{Workman:2012jf,Chen:2012yv,Workman:2011vb}.
Contradictory results for the $\gamma n$ coupling $A^{n}_{1/2}$ of
$N(1650)S_{11}$ obtained from two different reactions with the same
model indicate that the $N(1650)S_{11}$ state found in the $\gamma
n\to \pi^0 n$ is possibly not the same state found in the $\gamma
n\to \eta n$ if the data are accurate enough. It should be noted
that the FSI is rather rough correction that assumes identical
effects on the proton and the neutron, which certainly does not have
to be the case~\cite{Dieterle:2014blj}. Thus, considering that the
data from the A2 Collaboration might bear large uncertainties in the
second resonance region, with a small positive helicity amplitude,
$A^{n}_{1/2}\simeq 20\times10^{-3}$GeV$^{-1/2}$ for $N(1650)S_{11}$,
we predict the differential and total cross sections around the
second resonance region ( see Fig.~\ref{fig-eff}). If
$N(1650)S_{11}$ has a positive helicity amplitude, it is found that
i) the differential cross section and the total cross section around
the second resonance region should be significantly larger than the
present data; and ii) $N(1650)S_{11}$ has obviously constructive
interference with $N(1535)S_{11}$ and $N(1520)D_{13}$, which can be
tested in future experiments. It was pointed out in
Ref.~\cite{Boika:2014aha} that the positive $A^{n}_{1/2}$ would
imply $N(1650)S_{11}$ should have a large $s\bar{s}$ component in
its wave function. To clarify the sign problem of the $\gamma n$
coupling for $N(1650)S_{11}$, more accurate data are needed.

We find a large helicity amplitude for $\Delta(1620)S_{31}$, which
is about a factor 2 larger than the PDG average
value~\cite{Agashe:2014kda}, and $30\%$ larger than the recent
results from the BnGa~\cite{Anisovich:2009zy,Anisovich:2013jya} and
SAID~\cite{Workman:2011vb} groups. However, we find that our result
is very close to the theoretical predictions in quark
models~\cite{Li:1990qu,Capstick:1992uc}.

In our previous work~\cite{Zhong:2011ti}, we gave our estimations of
the helicity amplitudes for $N(1520)D_{13}$ by the analysis of the
$\eta$ photoproduction data. However, the large uncertainties of the
branching ratio $b_{\eta N}$ lead to a weak conclusion of these
helicity amplitudes. In this work, the accurate branching ratio
$b_{\pi N}$ should let us extract the helicity amplitudes for
$N(1520)D_{13}$ more reliably. It is found that the $A^p_{1/2}$
extracted by us is in good agreement with the results from SAID
group~\cite{Workman:2011vb} and the PDG average
value~\cite{Agashe:2014kda}. However, the $A^p_{3/2}$ extracted in
present work are about 30\% smaller than the PDG average
value~\cite{Agashe:2014kda} and the results from other groups. It
should be mentioned that recently the CBELSA/TAPS Collaboration also
found a small helicity amplitude $A^p_{3/2}\simeq
118\times10^{-3}$GeV$^{-1/2}$ from an energy-independent multipole
analysis based on new polarization data on photoproduction of
neutral pions~\cite{Hartmann:2014mya}. The $\gamma n$ couplings for
the $N(1520)D_{13}$ extracted in this work are compatible with the
PDG values within 30\% uncertainties. Our results are slightly
smaller than the results from other partial wave analysis groups.

For $N(1720)P_{13}$, we have noted that the absolute values of the
$A^p_{1/2}$ and $A^p_{3/2}$ extracted by us are compatible with the
results from the BnGa~\cite{Anisovich:2009zy,Anisovich:2013jya} and
Kent~\cite{Shrestha:2012ep} groups. However, their solutions have
opposite signs to our results. It is interesting to find that our
results are consistent with the quark model predictions by Z. Li and
F. Close~\cite{Li:1990qu}, and the partial wave analysis of the
$\gamma n\to \eta n$ reaction from the Giessen
group~\cite{Shklyar:2012js}. Knowledge about the $\gamma n$
couplings, $A^n_{1/2}$ and $A^n_{3/2}$, for the $N(1720)P_{13}$ is
very poor, and different groups have given very different
predictions. In the SU(6)$\otimes$O(3) symmetry limit, we predict
the $A^n_{3/2}$ should be zero, which is compatible with the
analysis of the Kent group~\cite{Shrestha:2012ep}. More studies are
needed to clarify these puzzles about the $N(1720)P_{13}$.

\section{Summary}\label{sum}

In this work, we have studied neutral pion photoproduction on
nucleons within a chiral quark model. We have achieved reasonable
descriptions of the data from the pion production threshold up to
the second resonance region.

The roles of the low-lying resonances in the reactions were
carefully analyzed. We found that: (i) $\Delta(1232)P_{33}$,
$N(1535)S_{11}$, $N(1520)D_{13}$, and $N(1720)P_{13}$ play crucial
roles in both $\gamma p\to \pi^0p$ and $\gamma n\to \pi^0n$
reactions. The $\Delta(1232)P_{33}$ resonance not only plays a
dominant role around the first resonance region, but also
contributes up to the third resonance region. Both $N(1535)S_{11}$
and $N(1520)D_{13}$ paly crucial roles around the second resonance
region. The second bump structure around $E_\gamma=700$ MeV in the
cross section receives approximately equal contributions from these
two resonances. $N(1720)P_{13}$ might play a crucial role in the
third resonance region. It might be responsible for the third bump
structure in cross section, which should be further investigated due
to our relatively poor descriptions of the polarization observables
in this energy region. (ii) Furthermore, obvious evidence of
$N(1650)S_{11}$ and $\Delta(1620)S_{31}$ is also found in the
reactions. They notably affect the cross sections and the
polarization observables from the second resonance region to the
third resonance region. (iii) The $u$- and $t$-channel backgrounds
play a crucial role in the reaction as well. The $u$ channel has
strong interferences with the resonances, such as
$\Delta(1232)P_{33}$, $N(1535)S_{11}$ and $N(1520)D_{13}$. By
including the $t$-channel vector-meson exchange contribution, the
descriptions of the data in the energy region $E_\gamma= 600\sim
900$ MeV are improved notably. (iv) No obvious evidence of the other
resonances, e.g., $N(1700)D_{13}$, $N(1675)D_{15}$,
$\Delta(1700)D_{33}$ and $N(1680)F_{15}$, was found in the
reactions.

Furthermore, the helicity couplings for the main resonances,
$\Delta(1232)P_{33}$, $N(1535)S_{11}$, $N(1520)D_{13}$,
$N(1720)P_{13}$, $N(1650)S_{11}$ and $\Delta(1620)S_{31}$, were
extracted from the reactions. We found that: (i) Our extracted
helicity amplitudes of $\Delta(1232)P_{33}$ and $N(1535)S_{11}$ are
in good agreement with the PDG average values and the results of
other groups. (ii) The $\gamma p$ coupling for $N(1650)S_{11}$
extracted by us is in good agreement with the results from the
SAID~\cite{Workman:2011vb, Workman:2012jf,Chen:2012yv} and
BnGa~\cite{Anisovich:2009zy,Anisovich:2013jya}. However, properties
of the $\gamma n$ coupling for $N(1650)S_{11}$ are still
controversial. Our analysis of the recent data of the $\gamma n\to
\pi^0 n$ reaction indicates a small negative $\gamma n$ coupling for
$N(1650)S_{11}$. Its sign is opposite to that of other analyses of
the $\gamma n\to \eta n$ data~\cite{Zhong:2011ti,
Anisovich:2013jya,Anisovich:2015tla}. (iii) We obtain a large
helicity coupling for $\Delta(1620)S_{31}$, but it is very close to
the recent analysis from the BnGa
group~\cite{Anisovich:2009zy,Anisovich:2013jya}. (iv) We give
smaller helicity couplings for $N(1520)D_{13}$, which are compatible
with the PDG values at the 30\% level. (v) The helicity couplings
$A^p_{1/2}$ and $A^p_{3/2}$ for $N(1720)P_{13}$ extracted by us are
consistent with the quark model predictions by Li and
Close~\cite{Li:1990qu,Capstick:1992uc} and the analysis of the
Giessen group~\cite{Shklyar:2012js}. We find a small positive
helicity coupling $A^n_{1/2}$ for $N(1720)P_{13}$, and the
$A^n_{3/2}$ should be zero in the SU(6)$\otimes$O(3) symmetry limit.

Finally, it should be pointed out that (i) the width of
$N(1720)P_{13}$ extracted by us is notably narrower than the
estimated values from the PDG, however, our result is in good
agreement with those extracted from the $\pi^-p\to K^0 \Lambda$
reaction by Saxon {\it et al.}~\cite{Saxon:1979xu}. To confirm the
properties of $N(1720)P_{13}$, a study of the $\pi^-p\to K^0
\Lambda$ reaction is needed. (ii) Furthermore, a more realistic
correction of the FSI for neutral pion photoproduction on quasi-free
neutrons hopefully will be obtained in future. Then the sign problem
of the $\gamma n$ coupling $A^{n}_{1/2}$ of $N(1650)S_{11}$ could be
clarified in the $\gamma n\rightarrow \pi^0 n$ reaction, which seems
to be crucial to uncover the puzzle of the narrow structure around
$W=1.68$ GeV observed in the excitation function of $\eta$
production off quasi-free neutrons. If the $\gamma n$ coupling
$A^{n}_{1/2}$ of the $N(1650)S_{11}$ is negative, then the narrow
structure in the $\gamma n\to \eta n$ reaction would no longer be
explained by the interference effects between $N(1535)S_{11}$ and
$N(1650)S_{11}$.

\section*{  Acknowledgements }

The authors would like to thank Qiang Zhao for useful discussions.
We also thank B. Krusche and M. Dieterle for providing us the data
of neutral pion photoproduction on the nucleons, Jan Hartmann for
providing us the new data of polarization observables $T$, $P$ and
$H$ for the $\gamma p\rightarrow \pi^0 p$ reaction, Paolo Levi
Sandri for providing us the data on beam asymmetry for the $\gamma
n\rightarrow \pi^0 n$ reaction. This work is supported, in part, by
the National Natural Science Foundation of China (Grants No.
11075051, No. 11375061, and No. 11405222), and the Hunan Provincial
Natural Science Foundation (Grant No. 13JJ1018).



\end{document}